\documentclass[fleqn,usenatbib]{mnras}
\usepackage{newtxtext,newtxmath}
\usepackage{graphics,graphicx}	% Including figure files
\usepackage{amsmath}	% Advanced maths commands
\usepackage{float}
\usepackage[T1]{fontenc}
\usepackage{ae,aecompl}
\usepackage{comment}

\newcommand{\fluxcgs}{erg~cm$^{-2}$~s$^{-1}$}

\title[X-ray polarimetry of the Circinus galaxy]{Mapping the circumnuclear regions of the Circinus galaxy with the Imaging X-ray Polarimetry Explorer}

      \author[]{F. Ursini,$^{\ref{aff:UniRoma3}}$\thanks{E-mail: francesco.ursini@uniroma3.it} A. Marinucci,$^{\ref{aff:ASI}}$ G. Matt,$^{\ref{aff:UniRoma3}}$ S. Bianchi,$^{\ref{aff:UniRoma3}}$ F. Marin,$^{\ref{aff:Strasbourg}}$ H.~L. Marshall,$^{\ref{aff:MIT}}$
      R. Middei,$^{\ref{aff:ASI}}$ 
              \newauthor
    J. Poutanen,$^{\ref{aff:Turku},\ref{aff:IKI}}$
    D. Rogantini,$^{\ref{aff:MIT}}$
    A. De Rosa,$^{\ref{aff:INAF-IAPS}}$ L. Di Gesu,$^{\ref{aff:ASI}}$ J.~A. Garc\'ia,$^{\ref{aff:Caltech}}$ A. Ingram,$^{\ref{aff:Oxford},\ref{aff:Newcastle}}$
    D.~E. Kim,$^{\ref{aff:INAF-IAPS},\ref{aff:UniRoma1},\ref{aff:UniRoma2}}$ 
            \newauthor
    H. Krawczynski,$^{\ref{aff:WUStL}}$
    S. Puccetti,$^{\ref{aff:ASI}}$ P. Soffitta,$^{\ref{aff:INAF-IAPS}}$ J. Svoboda,$^{\ref{aff:CAS-ASU}}$ 
    F. Tombesi,$^{\ref{aff:UniRoma2}, \ref{aff:INFN-Roma2}, \ref{aff:UMd}}$
    M.~C. Weisskopf,$^{\ref{aff:NASA-MSFC}}$
        \newauthor
    T. Barnouin,$^{\ref{aff:Strasbourg}}$         M.~Perri,$^{\ref{aff:ASI-SSDC},\ref{aff:INAF-OAR}}$ J. Podgorny,$^{\ref{aff:Strasbourg},\ref{aff:CAS-ASU},\ref{aff:Charles}}$ 
        A. Ratheesh,$^{\ref{aff:INAF-IAPS}}$ 
        A. Zaino,$^{\ref{aff:UniRoma3}}$
        I. Agudo,$^{\ref{aff:CSIC-IAA}}$
        L.~A.~Antonelli,$^{\ref{aff:INAF-OAR},\ref{aff:ASI-SSDC}}$
        \newauthor
        M.~Bachetti,$^{\ref{aff:INAF-OAC}}$
        L.~Baldini,$^{\ref{aff:INFN-PI},\ref{aff:UniPI}}$
        W.~H.~Baumgartner,$^{\ref{aff:NASA-MSFC}}$
        R.~Bellazzini,$^{\ref{aff:INFN-PI}}$
        S.~D.~Bongiorno,$^{\ref{aff:NASA-MSFC}}$
        R.~Bonino,$^{\ref{aff:INFN-TO},\ref{aff:UniTO}}$
        \newauthor
        A.~Brez,$^{\ref{aff:INFN-PI}}$
        N.~Bucciantini,$^{\ref{aff:INAF-FI},\ref{aff:UniFI},\ref{aff:INFN-FI}}$
        F.~Capitanio,$^{\ref{aff:INAF-IAPS}}$
        S.~Castellano,$^{\ref{aff:INFN-PI}}$
        E.~Cavazzuti,$^{\ref{aff:ASI}}$
        S.~Ciprini,$^{\ref{aff:INFN-Roma2},\ref{aff:ASI-SSDC}}$
        E.~Costa,$^{\ref{aff:INAF-IAPS}}$
        \newauthor
        E.~Del~Monte,$^{\ref{aff:INAF-IAPS}}$
        N.~Di~Lalla,$^{\ref{aff:Stanford}}$
        A.~Di~Marco,$^{\ref{aff:INAF-IAPS}}$
        I.~Donnarumma,$^{\ref{aff:ASI}}$
        V.~Doroshenko,$^{\ref{aff:Tuebingen}}$
        M.~Dovčiak,$^{\ref{aff:CAS-ASU}}$
        \newauthor
        S.~R.~Ehlert,$^{\ref{aff:NASA-MSFC}}$
        T.~Enoto,$^{\ref{aff:RIKEN}}$
        Y.~Evangelista,$^{\ref{aff:INAF-IAPS}}$
        S.~Fabiani,$^{\ref{aff:INAF-IAPS}}$
        R.~Ferrazzoli,$^{\ref{aff:INAF-IAPS}}$
        S.~Gunji,$^{\ref{aff:yamagata}}$
        J.~Heyl,$^{\ref{aff:UBC}}$
        W.~Iwakiri,$^{\ref{aff:Chuo}}$
        \newauthor
        S.~G.~Jorstad,$^{\ref{aff:BU},\ref{aff:SBU}}$
        V.~Karas,$^{\ref{aff:CAS-ASU}}$
        T.~Kitaguchi,$^{\ref{aff:RIKEN}}$
        J.~J.~Kolodziejczak,$^{\ref{aff:NASA-MSFC}}$
        F.~La~Monaca,$^{\ref{aff:INAF-IAPS}}$
        L.~Latronico,$^{\ref{aff:INFN-TO}}$
        \newauthor
        I.~Liodakis,$^{\ref{aff:FINCA}}$
        S.~Maldera,$^{\ref{aff:INFN-TO}}$
        A.~Manfreda,$^{\ref{aff:INFN-PI}}$
        A.~P.~Marscher,$^{\ref{aff:BU}}$
        I.~Mitsuishi,$^{\ref{aff:nagoya}}$
        T.~Mizuno,$^{\ref{aff:hiroshima}}$
        F.~Muleri,$^{\ref{aff:INAF-IAPS}}$
        \newauthor
        C.~Y.~Ng,$^{\ref{aff:HKU}}$
        S.~L.~O'Dell,$^{\ref{aff:NASA-MSFC}}$
        N.~Omodei,$^{\ref{aff:Stanford}}$
        C.~Oppedisano,$^{\ref{aff:INFN-TO}}$
        A.~Papitto,$^{\ref{aff:INAF-OAR}}$
        G.~G.~Pavlov, $^{\ref{aff:PSU}}$
        A.~L.~Peirson,$^{\ref{aff:Stanford}}$
        \newauthor
        M.~Pesce-Rollins,$^{\ref{aff:INFN-PI}}$
        P.-O.~Petrucci, $^{\ref{aff:Grenoble}}$
        M. Pilia,$^{\ref{aff:INAF-OAC}}$
        A. Possenti,$^{\ref{aff:INAF-OAC}}$
        B.~D.~Ramsey,$^{\ref{aff:NASA-MSFC}}$
        J.~Rankin,$^{\ref{aff:INAF-IAPS}}$
        \newauthor
        R.~W.~Romani,$^{\ref{aff:Stanford}}$
        C.~Sgrò,$^{\ref{aff:INFN-PI}}$
        P.~Slane,$^{\ref{aff:CfA}}$
        G.~Spandre,$^{\ref{aff:INFN-PI}}$
        T.~Tamagawa,$^{\ref{aff:RIKEN}}$
        F.~Tavecchio,$^{\ref{aff:INAF-OAB}}$
        R.~Taverna,$^{\ref{aff:UniPD}}$
        \newauthor
        Y.~Tawara,$^{\ref{aff:nagoya}}$
        A.~F.~Tennant,$^{\ref{aff:NASA-MSFC}}$
        N.~E.~Thomas,$^{\ref{aff:NASA-MSFC}}$
        A.~Trois,$^{\ref{aff:INAF-OAC}}$
        S.~S.~Tsygankov,$^{\ref{aff:Turku},\ref{aff:IKI}}$
        R.~Turolla,$^{\ref{aff:UniPD},\ref{aff:MSSL}}$
        J.~Vink,$^{\ref{aff:amsterdam}}$
        \newauthor
        K.~Wu,$^{\ref{aff:MSSL}}$
        F.~Xie,$^{\ref{aff:GXU},\ref{aff:INAF-IAPS}}$
        S.~Zane$^{\ref{aff:MSSL}}$
        \\
\\
Affiliations are shown at the end of the paper
}

\date{Accepted XXX. Received YYY; in original form ZZZ}

\pubyear{2022}

\begin{document}
\label{firstpage}
\pagerange{\pageref{firstpage}--\pageref{lastpage}}
\maketitle
\begin{abstract}
 We report on the Imaging X-ray Polarimetry Explorer (\textit{IXPE}) observation of the closest and X-ray brightest Compton-thick active galactic nucleus (AGN), the Circinus galaxy. We find the source to be significantly polarized in the 2--6 keV band. From previous studies, the X-ray spectrum is known to be dominated by reflection components, both neutral (torus) and ionized (ionization cones). Our analysis indicates that the polarization degree is $28\pm7$ per cent (at 68 per cent confidence level) for the neutral reflector, with a polarization angle of $18\degr\pm5\degr$, roughly perpendicular to the radio jet.
The polarization of the ionized reflection is unconstrained.
 A comparison with Monte Carlo simulations of the polarization expected from the torus shows that the neutral reflector is consistent with being an equatorial torus with a half-opening angle of 45\degr--55\degr.
  This is the first X-ray polarization detection in a Seyfert galaxy, demonstrating the power of X-ray polarimetry in probing the geometry of the circumnuclear regions of AGNs, and confirming the basic predictions of standard Unification Models.
\end{abstract}


%%%%%%%%%%%%%%%%%%%%%%%%%%%%%%%%%%%%%%%%%%%%%%%%%%

%%%%%%%%%%%%%%%%% BODY OF PAPER %%%%%%%%%%%%%%%%%%

\section{Introduction}

A large fraction of active galactic nuclei (AGNs) are obscured by gas and dust \citep[$\sim 70$ per cent of local AGNs; e.g.][]{ramos-almeida&ricci2017}. According to Unification Models \citep[e.g.][]{antonucci1993}, the obscuring medium is a geometrically thick and axisymmetric structure, and is generally referred to as the ``torus''. Sources obscured by material with a column density $N_{\rm H}> \sigma_{\rm T}^{-1} = 1.5 \times 10^{24}$ cm$^{-2}$ (where $\sigma_{\rm T}$ is the Thomson cross-section) are called Compton-thick, and make up a sizeable fraction of local AGNs \citep[$\sim 20{-}30$ per cent; e.g.][]{malizia2009,ricci2015,torres-alba2021}.
In general, the X-ray nuclear radiation from Compton-thick Seyfert 2 galaxies is completely obscured, at least up to 10 keV \citep[e.g.][]{arevalo2014,marinucci2016}. The X-ray spectrum is thus dominated by reflection, from both neutral and ionized matter surrounding the nucleus \citep[e.g.][]{matt2000}.
%This is not the case in unobscured objects, in which reflection is often heavily diluted, or even invisible.
For unobscured objects, the direct  emission dominates and
the reflection component is weak or even invisible.
%Most of our knowledge on the circumnuclear matter, at least as far as their X--ray properties are concerned, are based on the brightest Compton--thick sources, like NGC 1068 (Kinkhabwala et al. 2002; Matt et al. 2004; Bauer et al. 2015; Marinucci et al. 2016; Zaino et al. 2020), Circinus (Matt et al. 1996; Ar{\'e}valo et al. 2014), NGC 4945 (Puccetti et al. 2014, Marinucci et al. 2012, 2017), Mrk 3 (Pounds $\&$ Page 2005; Guainazzi et al. 2012, 2016), Mrk 573 (Bianchi et al. 2010, Paggi et al. 2012) and ESO 428-G014 (Fabbiano et al. 2017, Feruglio et al. 2020).
Compton-thick AGN are therefore ideal candidates for X-ray polarimetric observations aimed at determining the geometry of the circumnuclear scattering material. In fact, the X-ray polarization produced by the reprocessing of the nuclear emission carries information about the cold reflector (i.e. the torus), responsible for the intense iron K$\alpha$ emission line and Compton reflection continuum, and the ionized reflector, responsible for the soft X-ray continuum and line emission. In particular, it has been shown that measurements of the half-opening angle of the torus and of the inclination angle with respect to the line of sight are possible, along with a comparison of the main axes of the X-ray reflecting structures with those of optical/IR emitting regions, like the ionization cone \citep[e.g.][]{goosmann&matt2011}. 

The Circinus galaxy is one of the closest AGN \citep[redshift $z=0.001449$; distance $D= 4.2\pm0.8$ Mpc,][]{freeman1977}.
It has been extensively observed by X-ray satellites in the last 30 years. It was detected by {\it ROSAT} during the All Sky Survey for the first time in the X-rays \citep{brinkmann1994} and later on by the {\it Advanced Satellite for Cosmology and Astrophysics} ({\it ASCA}), showing a spectrum dominated by a Compton reflection component \citep{matt1996}, with a prominent iron K$\alpha$ emission line and several other lines from lighter elements \citep{bianchi2001}. {\it Beppo-SAX} confirmed the {\it ASCA} results below 10 keV and the Compton-thick nature of the source at higher energies \citep{guainazzi1999,matt1999}. The properties of the nuclear and circumnuclear emission have then been investigated in more details thanks to the better angular and spectral resolution of {\it Chandra} \citep{sambruna2001_im,sambruna2001_hrsp,marinucci2013,kawamuro2019} and {\it XMM-Newton} \citep{molendi2003,massaro2006}. {\it NuSTAR}, more recently, confirmed the Compton-thick nature of the source \citep[$N_{\rm H}>6\times 10^{24}$ cm$^{-2}$,][]{arevalo2014}. 
The extranuclear activity of Circinus has been well studied also at longer wavelengths, showing a prominent [O\,{\sc iii}] ionization cone \citep{marconi1994}, two starburst rings at $\sim$2\arcsec\ and 10\arcsec\ from the nucleus \citep[][and references therein]{wilson2000}, and a radio jet has been observed as part of an overall complex extended radio structure \citep{elmouttie1998b,curran2008}. The 100-pc bipolar jet emanates from the compact core in the same direction of kpc-scale radio plumes \citep{elmouttie1998b}. On the subparsec scales, the source shows an edge-on, warped accretion disc, traced by H$_2$O maser emission at 1.3 cm with very long baseline interferometry \citep{greenhill2003}. 

The Circinus galaxy is the X-ray brightest \citep[$F_{\textrm{2--10\,keV}} \approx 1.5 \times 10^{-11}$ \fluxcgs,][]{bianchi2002} Compton-thick Seyfert 2 galaxy in the sky. Therefore, it has been chosen as the first Compton-thick AGN to be observed by the {\it Imaging X-ray Polarimetry Explorer} \citep[\textit{IXPE},][]{weisskopf2022,soffitta2021}. Here we report on this observation, performed jointly with \textit{Chandra} to enable a spatially resolved spectro-polarimetric study.

The paper is organized as follows. In Sect.~\ref{sec:data}, we describe the \textit{IXPE} and \textit{Chandra} observations and data reduction. In Sect.~\ref{sec:polar}, we report on the data analysis and results. In Sect.~\ref{sec:disc}, we discuss the constraints on the geometry of the system by comparing the results with detailed Monte Carlo simulations of the polarization expected from the torus.

%Among the 15 sources detected within a 1$'$ aperture radius of the nucleus (Bauer et al. 2001), the brightest ones are CG X-1 (CXOU J 141312.3-652013) and CG X-2 (CXOU J 141310.0-652044), at angular distances of 15$''$ north-east and 25$''$ south-west from the AGN, respectively (Fig. 1). Indeed, an anomalous change in flux and spectral shape was reported for a {\it Beppo}SAX observation in 2001, then interpreted as an increase in brightness of the ULX CG X-1 (Bianchi et al. 2002). CG X-1 is an eclipsing X-ray binary system with a Wolf-Rayet donor star, a period P$\simeq$7.2 hr and a 2-8 keV flux which varies from a few percent to $\sim45\%$ of the one of the AGN on timescales of weeks (Esposito et al. 2015, Qiu et al. 2019). CG X-2, on the other hand, is a young supernova remnant candidate (SN 1996cr: Bauer et al. 2008) with a stable flux of a few percent with respect to the nuclear one. 

\section{Observations and data reduction}
\label{sec:data}
\subsection{\textit{IXPE}}
{\it IXPE} observed the Circinus galaxy starting on 2022 July 12 with its three Detector Units (DU)/Mirror Module Assemblies (MMA), for a net exposure time of 771.5 ks (Table \ref{tab:obs-log}). Cleaned level 2 event files were produced and calibrated using standard filtering criteria with the dedicated {\sc ftools} tasks\footnote{\url{https://heasarc.gsfc.nasa.gov/docs/ixpe/analysis/IXPE-SOC-DOC-009-UserGuide-Software.pdf}} and the latest calibration files available in the {\it IXPE} calibration database (CALDB 20220314). 

The Stokes $I$, $Q$, $U$ background spectra were extracted from source-free circular regions with a radius of 100\arcsec. Extraction radii for the $I$ Stokes spectra of the source were computed via an iterative process which leads to the maximization of the signal-to-noise ratio (SNR) in the 2--8 keV energy band, similar to the approach described in \citet{piconcelli2004}. This procedure consists in the extraction of several source spectra, monotonically increasing the value of the radius, until the maximum SNR is found. We eventually adopted circular regions centered on the source with radii of 52\arcsec, 52\arcsec\ and 47\arcsec\ for DU1, DU2 and DU3, respectively. 
%The net exposure times are 770\,ks and 
The same extraction radii were then applied to the $Q$ and $U$ Stokes spectra. The weighted analysis method presented in \citet{dimarco22} (parameter {\tt stokes=Neff} in {\sc xselect}) was applied. We used a constant energy binning of 0.2 keV for $Q$, $U$ Stokes spectra and required a 2--8 keV SNR higher than 3 in each spectral channel, in the flux spectra. $I$, $Q$, $U$ Stokes spectra from the three DU/MMAs are always fitted independently in the following. Background represents 17.0, 19.7 and 16.4 per cent of the total DU1/MMA1, DU2/MMA2 and DU3/MMA3 $I$ flux spectra, respectively. 
The same extraction regions were used in {\sc ixpeobssim} \citep[version 28.2.0:][]{baldini22}, which implements the method of \cite{kislat2015} to reconstruct the polarization properties. The information is represented as data cubes containing images of the Stokes parameters, binned in sky coordinates. Polarization cubes from the source and from the background were extracted, following the same procedure as for MCG-05-23-16 \citep{marinucci2022}, including background subtraction. 
We created one polarization cube for each DU in the 2--8 keV band and then combined them; we then computed the Stokes parameters, the corresponding polarization degree and angle, and the associated uncertainties. 

\subsection{\textit{Chandra}}
\label{sec:chandra}
The central region of the Circinus galaxy is heavily populated by ultraluminous X-ray (ULX) sources \citep{bauer2001,smith&wilson2001}, and the \textit{IXPE} point spread function ($\simeq30\arcsec$) does not allow us to spatially resolve some of them. For this reason, {\it Chandra} observed the Circinus galaxy at the beginning and end of the {\it IXPE} observation. These two {\it Chandra} observations allowed us to monitor the flux state of the ULXs and in particular of CG X-1, which is highly variable, reaching in the past fluxes almost as large as that of the AGN \citep{bianchi2002}. 

The two \textit{Chandra} observations were performed on 2022 July 11 and July 24 (see Table \ref{tab:obs-log}) with the Advanced CCD Imaging Spectrometer \citep[ACIS;][]{acis} for elapsed times of 10\,ks each. To reduce pileup effects, the frame time was set to 0.5\,s and custom CCD subarrays were used. Data were reduced with the Chandra Interactive Analysis of Observations \citep[\textsc{CIAO};][]{ciao} 4.14 and the Chandra Calibration Data Base 4.9.8, adopting standard procedures. At the distance of the source 1\arcsec\ corresponds to 19~pc. We generated event files for the two observations with the {\sc CIAO} tool {\tt chandra$_{-}$repro} and, after cleaning for background flaring, we got net exposure times of 9.3\,ks each.

\begin{figure}
\centering
\includegraphics[width=0.8\columnwidth]{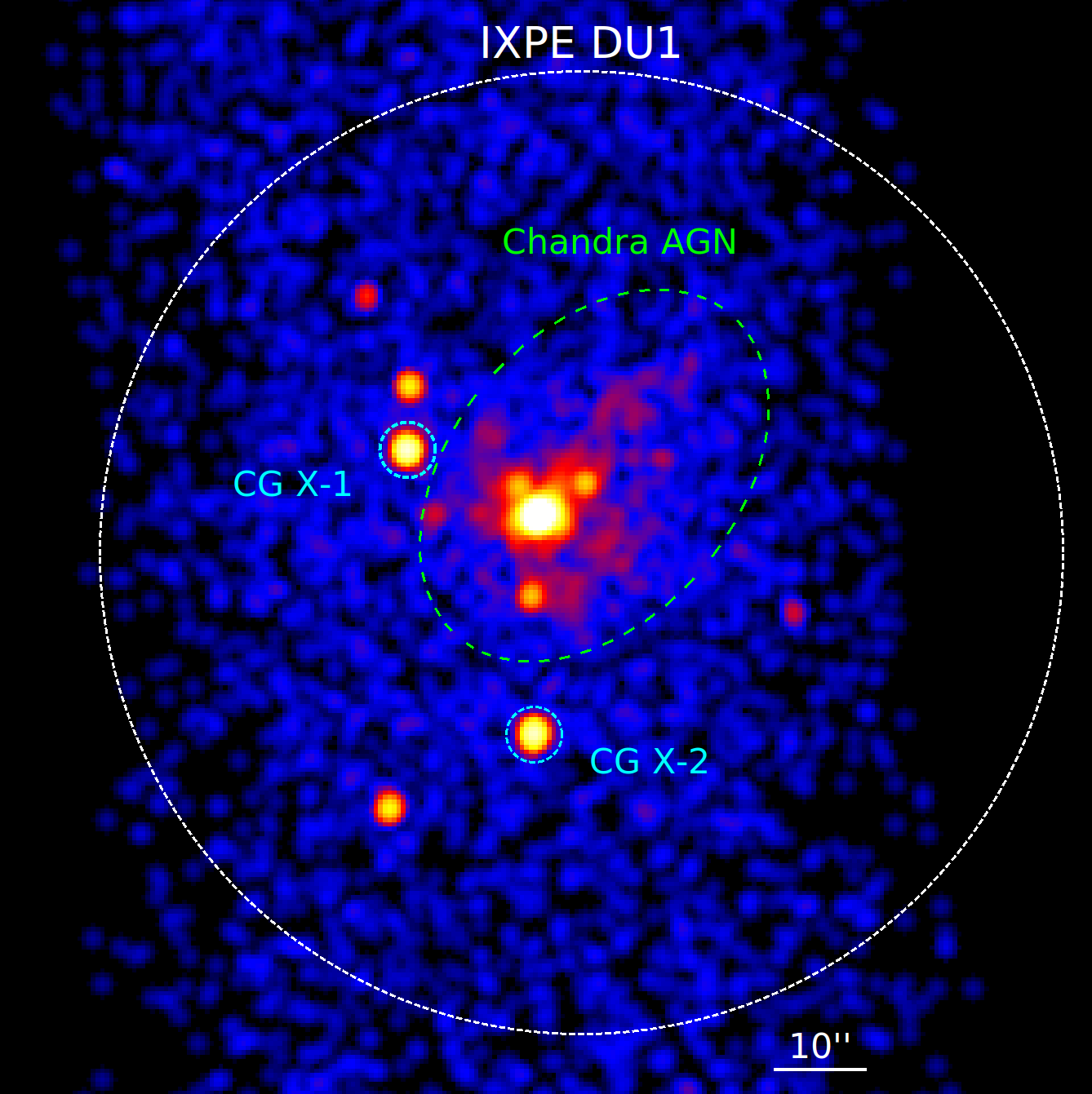}
\caption{\textit{Chandra} image of the Circinus galaxy (2\arcmin $\times$ 2\arcmin). The green dashed ellipse marks the \textit{Chandra} extraction region for the AGN (see Sect.~\ref{sec:chandra} for the details). The cyan dotted lines encircle the ULXs CG X-1 and CG X-2. The white dotted circle corresponds to the \textit{IXPE} DU1 extraction region.
  }
\label{fig:chandra-image}
\end{figure}

The spectra from the CG~X-1 and CG~X-2 were extracted from circular regions centered on the two sources with 3\arcsec\ radii. We used a 10\arcsec\ radius circle for background extraction. For the AGN, we used an ellipse with a semi-minor axis of 15\arcsec\ and semi-major axis of 23\arcsec\ (Fig. \ref{fig:chandra-image}). 
Co-adding the two exposures, in the 2--8 keV band 800 counts are measured for the two ULXs and 4910 counts for the AGN.

After the two ULXs, the brightest point sources in the field, both outside of the AGN extraction region, are CXOU~J141312.6--652052 and CXOU~J141312.2--652007 \citep{bauer2001}. However, only 210 counts (2--8 keV) are measured in total for the two sources together. Within the AGN extraction region, three other point sources are detected, namely CXOU~J141309.2--652017, CXOU~J141310.1--652029, and CXOU~J141310.3--652017 \citep{bauer2001}. Evaluating their counts is less obvious due to the contamination of the AGN, however we estimate a total of 250 counts (2--8 keV) summing all the three sources. Given the overall small contribution of these sources, for simplicity we only take into account the contribution from CG~X-1 and CG~X-2 in the following spectral analysis. 
%This choice does not significantly affect the polarization results.

Spectra were binned in order to oversample the instrumental resolution by a factor of 3 and to have no less than 30 counts in each background-subtracted spectral channel. This allows the applicability of the $\chi^2$ statistic. We ignored channels between 8 and 10 keV due to pileup, which is not unexpected owing to the decreased effective area of the detector and the decline of the intrinsic source spectrum.\footnote{For more details, see the Chandra ABC Guide to Pileup, online at: https://cxc.harvard.edu/ciao/download/doc/pileup\_abc.pdf.} We estimate the fraction of detected events that are in fact piled using the {\sc ciao} tool {\tt pileup$_{-}$map}.
We infer that in the central $3\times3$ pixels region the average pileup fraction is 6 per cent, ranging from 3 to 9 per cent (in the central pixel) for both observations. 
Details on the analysis of {\it Chandra} images and spectra are reported in the Appendix.  Since the AGN emission is consistent with being constant (see Appendix), we use the co-added spectrum for the spectro-polarimetric analysis.

\begin{table}
\begin{center}
  \begin{tabular}{cccc}
    \hline    \hline
 Satellite & Obs. Id. & Start time (UTC) & Net exp. (ks) \\
\hline 
\textit{IXPE} & 01003501 & 2022-07-12 & 771.5 \\
\textit{Chandra} & 25365 & 2022-07-11 & 9.3 \\
 & 25366 & 2022-07-24 & 9.3 \\
\hline 
\end{tabular}
\end{center}
\caption{\label{tab:obs-log}
Logs of the \textit{IXPE} and \textit{Chandra} observations.
}
\end{table}

\begin{table}
\begin{center}
  \begin{tabular}{ccc}
    \hline    \hline
 Energy & P.D. (\%) & P.A. (deg) \\
\hline 
2--8 keV&  $17.6\pm3.2$ &  $16.9\pm5.3$  \\
2--4 keV &$16.0\pm4.9$ & $19.1\pm8.9$ \\
4--6 keV&$26.3\pm5.7$ & $20.2\pm7.5$  \\
2--6 keV &$20.0\pm3.8$ & $19.1\pm5.5$ \\
6--8 keV &$<24.5$ & - \\
\hline 
\end{tabular}
\end{center}
\caption{\label{tab:poldeg-ang}
\textsc{xspec} measured polarization degrees and angles with associated uncertainties at 68 per cent ($1 \sigma$) confidence level for one parameter of interest.
For the 6--8 keV band, we report the upper limit at 99 per cent confidence level.
We note that the polarization angle is undefined when the polarization degree is not significantly detected.}
\end{table}

\begin{comment}
\begin{table}
\begin{center}
  \begin{tabular}{ccccc}
    \hline    \hline
 Energy & \multicolumn{2}{c}{P.D. (\%)} & \multicolumn{2}{c}{P.A. (deg)} \\
  & {\sc Pcubes} & {\sc Xspec} & {\sc Pcubes} & {\sc Xspec} \\
\hline 
2--8 keV& $15.3\pm4.8$  & $17.6\pm3.2$ & $10.3\pm8.9$& $16.9\pm5.3$  \\
2--4 keV& $19.3\pm5.3$ &$16.0\pm4.9$ & $19.9\pm7.8$ &$19.1\pm8.9$ \\
4--6 keV& $26.9\pm6.6$ &$26.3\pm5.7$ & $16.4\pm7.0$ &$20.2\pm7.5$  \\
2--6 keV& $22.8\pm4.4$ &$20.0\pm3.8$ & $18.0\pm5.5$ &$19.1\pm5.5$ \\
6--8 keV& $10.3\pm8.2$ &$8.9\pm6.3$ & - & - \\
\hline 
\end{tabular}
\end{center}
\caption{\label{tab:poldeg-ang}
Polarization degrees and angles with associated uncertainties (68 per cent confidence level). We note that the polarization angle is undefined when the polarization degree is not significantly detected.}
\end{table}
\end{comment}

\section{Polarization properties of the circumnuclear matter}
\label{sec:polar}

\begin{comment}
\begin{figure}
\includegraphics[width=1.0\columnwidth]{PD-PA-xspec.pdf}
\caption{Polarization degree (\textit{top panel}) and  polarization angle  (\textit{bottom panel}) as a function of energy, measured with 
%the polarization cubes (black) and with 
\textsc{xspec}. }
\label{poldeg-ang}
\end{figure}
\end{comment}

\begin{figure*}
\includegraphics[width=1.0\textwidth]{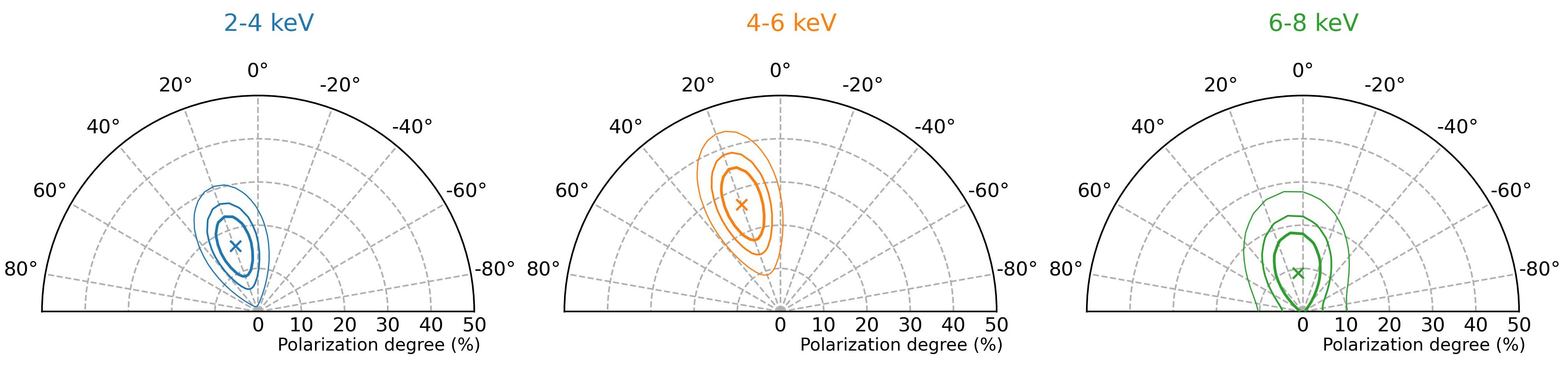}
\caption{Contour plots of the polarization degree and angle, measured with \textsc{xspec}, at the 68, 90 and 99 per cent confidence levels, in the  2--4  (left panel, light blue), 4--6  (middle panel, orange) and 6--8 keV (right panel, green) energy bands.}
\label{cplot}
\end{figure*}

\begin{figure*} 
\centering 
\includegraphics[width=0.9\columnwidth]{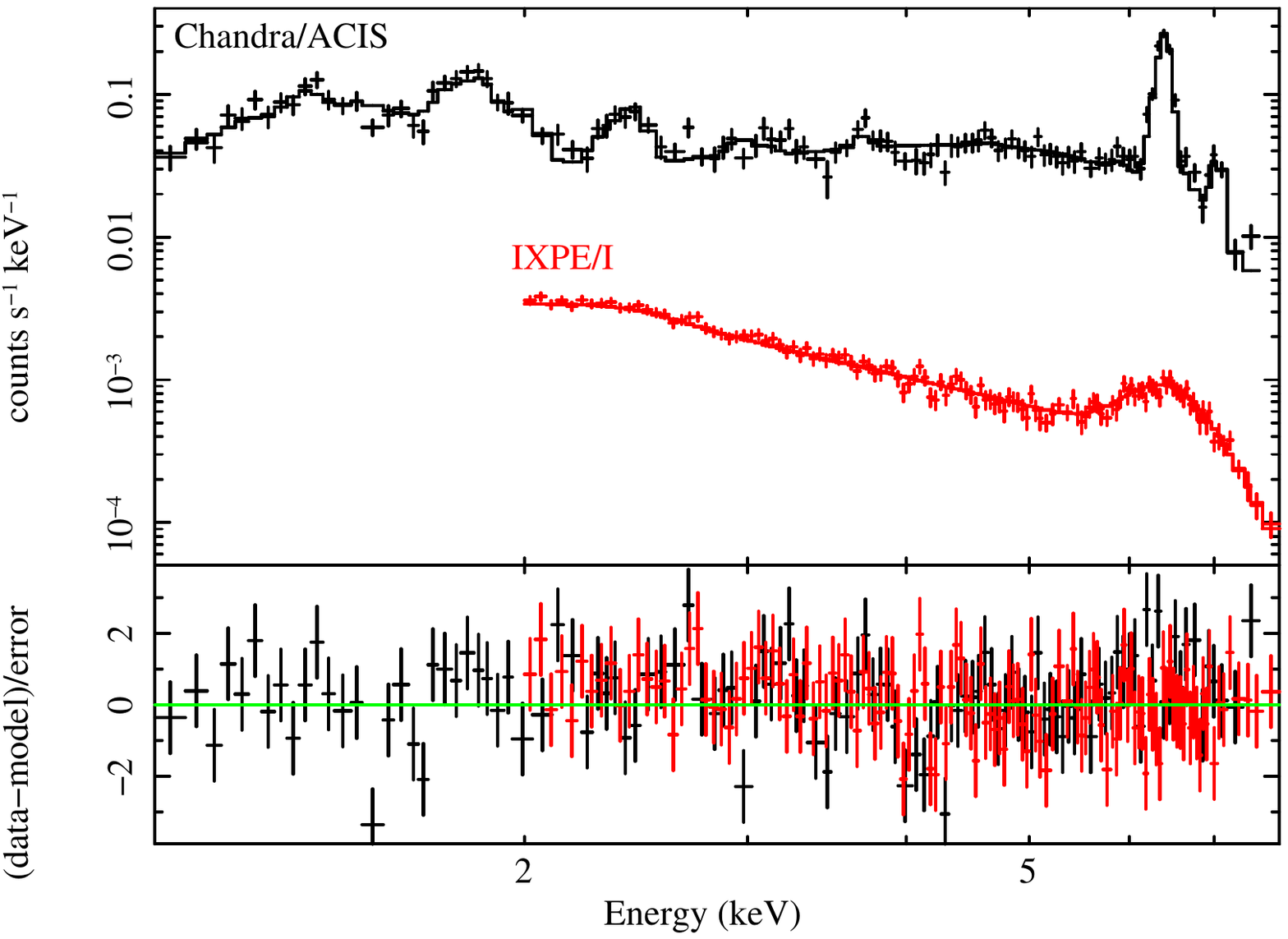}
\includegraphics[width=0.9\columnwidth]{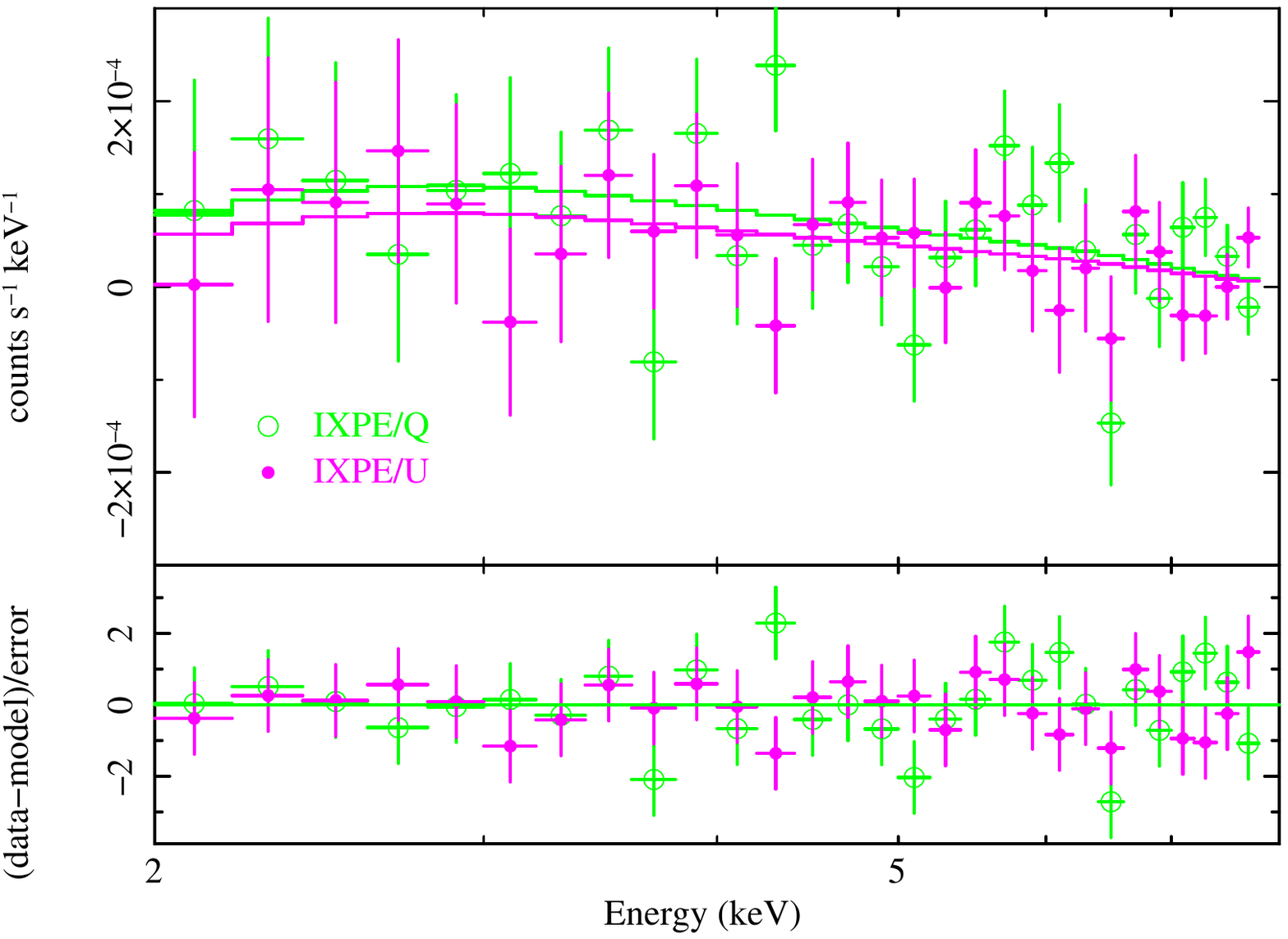}
\caption{\textit{Left panel:} \textit{Chandra}/ACIS and \textit{IXPE} $I$ (flux) spectra of the Circinus galaxy with the best-fitting model and the residuals. 
\textit{Right panel:} \textit{IXPE} $Q$ and $U$ Stokes spectra with the best-fitting model and the residuals. Note the different scale from the left panel. In both panels, \textit{IXPE} spectra are grouped for plotting purposes only (with \textsc{setplot group} in \textsc{xspec}).
}
\label{fig:bestfit}
\end{figure*}

In the {\it IXPE} energy range, the spectrum of the Circinus galaxy is known to be dominated by two components: emission from a warm reflector, likely due to ionized, optically thin matter and mostly contributing below 3 keV, and from a cold reflector dominating the spectrum above 4 keV, due to the optically thick torus \citep[e.g.][]{bianchi2001,sambruna2001_im,sambruna2001_hrsp,massaro2006}. The spectra of both reflectors are composed of a continuum (expected to be significantly polarized) and emission lines (expected to be
unpolarized, apart from the contribution by resonant scattering).
%In Fig.~\ref{poldeg-ang} 
In Table \ref{tab:poldeg-ang}
we report the polarization degree and angle at different energies. 
%(see also Table \ref{tab:poldeg-ang}). 
We show these parameters as measured from the \textit{IXPE} spectra using \textsc{xspec} 12.12.1 \citep{xspec}, with the 68 per cent confidence level uncertainty for one parameter of interest. The parameters found from the polarization cubes extracted with \textsc{ixpeobssim} are well consistent within the errors. We obtain a significant detection in the 2--6 keV band, with a polarization degree of $20.0\pm3.8$ per cent (at 68 per cent confidence level). 
The lack of a significant detection in the 6--8 keV band is likely due to the prominent iron lines present in the spectrum. 
Indeed, the flux of the iron lines between 6.4 and 7.057 keV reaches 50 per cent of the total flux in the 6--8 keV band \citep{massaro2006}. 
The ranges of polarization degree and angle shown in Table \ref{tab:poldeg-ang} are one-dimensional, in the sense that they are derived for each parameter independently of the other.
In Fig.~\ref{cplot}, we show two-dimensional contour plots of the polarization degree and angle at different energies. We note that %in the 6--8 keV band, we do not obtain a highly significant detection, while 
in the 2--4 and 4--6 keV bands the detection significance is greater than 99 per cent, and greater than $3\sigma$ at least in the 4--6 keV band. Finally, the polarization angle is consistent with being constant in energy.

\begin{table}
\begin{center}
  \begin{tabular}{ll}
\hline
\hline
Parameter & Value \\ 
\hline 
   \multicolumn{2}{c}{CG X-1 (\textsc{powerlaw})} \\
   $\Gamma$ & $1.1$(f) \\
    $N$ & $6 \times 10^{-5}$(f) \\
       \multicolumn{2}{c}{CG X-2 (\textsc{raymond+zgauss})} \\
   $kT_{\textsc{raymond}}$ (keV) & $10$(f) \\
    $N_{\textsc{raymond}}$ & $3 \times 10^{-4}$(f) \\
    $E_{\textsc{zgauss}}$ (keV) & $6.67$(f) \\
    $N_{\textsc{zgauss}}$ (keV) & $7.9 \times 10^{-6}$(f) \\
   \multicolumn{2}{c}{Cold reflector (\textsc{pexrav})} \\
$\Gamma$ & $1.6$(f) \\
$N$ & $(2.06 \pm 0.04) \times 10^{-2}$ \\
P.D. (\%) & $28 \pm 7$ \\
P.A. (deg) & $18 \pm 5$\\
 \multicolumn{2}{c}{Warm reflector (\textsc{powerlaw})} \\
 $\Gamma$ & $3.0$(f) \\
 $N$ & $6.6 \times 10^{-4}$(f) \\
P.D. (\%) &unconstrained\\ 
P.A. (deg) & $=18$ \\
\multicolumn{2}{c}{Cross-calibration constants} \\
$C_{\rm DU1-ACIS}$ & $0.809 \pm 0.014$ \\
$C_{\rm DU2-ACIS}$ & $0.723 \pm 0.013$ \\
$C_{\rm DU3-ACIS}$ & $0.677 \pm 0.012$ \\
\multicolumn{2}{c}{Observed flux} \\
$F_{\rm 2-8keV}$ & $(1.00 \pm 0.01) \times 10^{-11}$ \\
\hline
$\chi^2$/d.o.f. & $702/662$\\
\hline 
\end{tabular}
\end{center}
\caption{\label{tab:Fit}
Best-fitting parameters (68 per cent confidence level for one parameter of interest) of the joint \textit{Chandra} and \textit{IXPE} fit. Normalizations are in units of photons keV$^{-1}$ cm$^{-2}$ s$^{-1}$, while the flux is in unit of \fluxcgs. (f) denotes a fixed parameter.  
  }
\end{table}

\begin{figure}
\includegraphics[width=1.0\columnwidth]{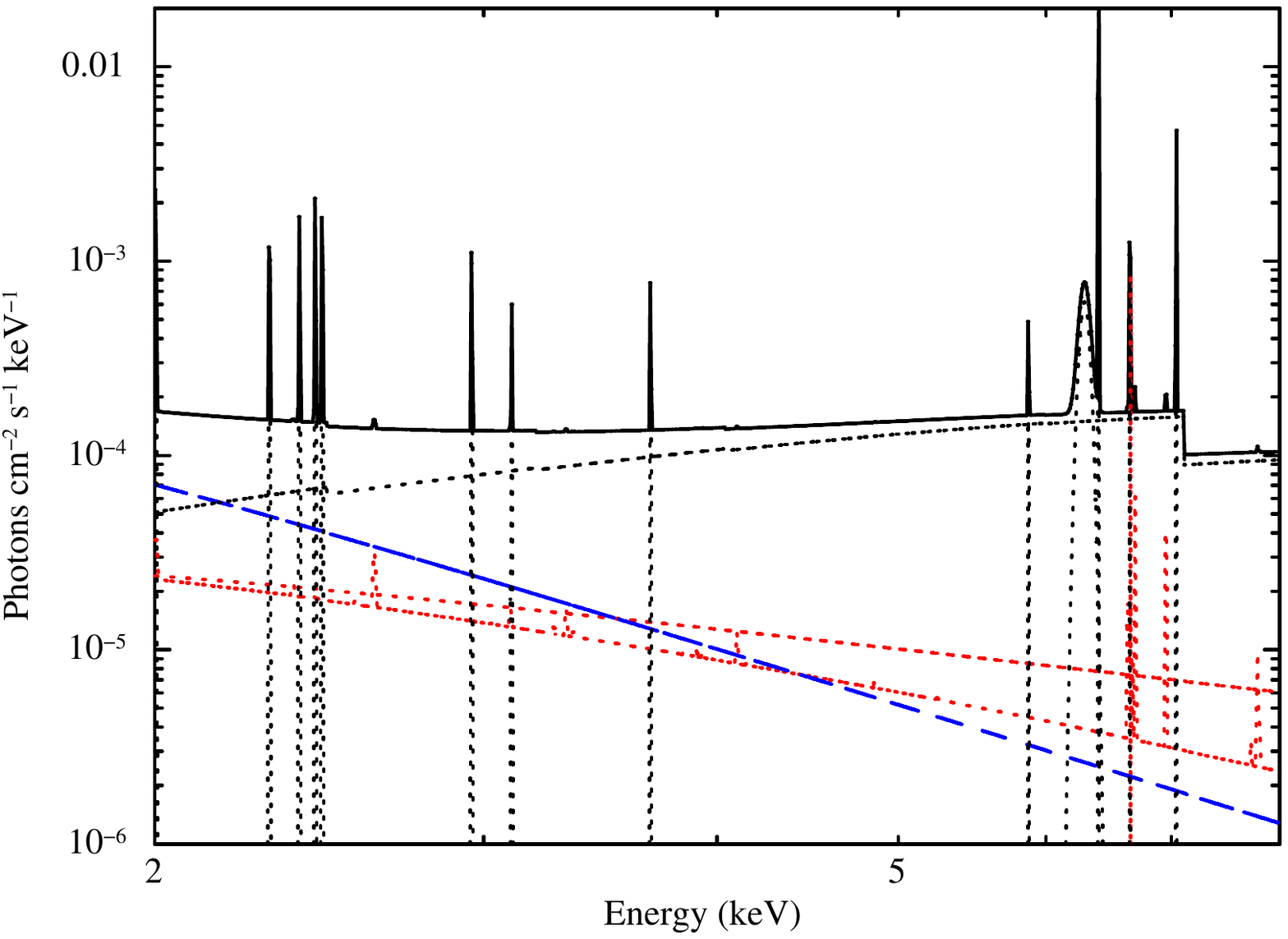}
\caption{Best-fitting total model (black solid line) in the 2--8 keV band, together with the contribution of various components: cold reflection (black dotted line), warm reflection (blue dashed line), ULXs (red dotted lines).}
\label{fig:model}
\end{figure}

\begin{figure}
\includegraphics[width=1.0\columnwidth]{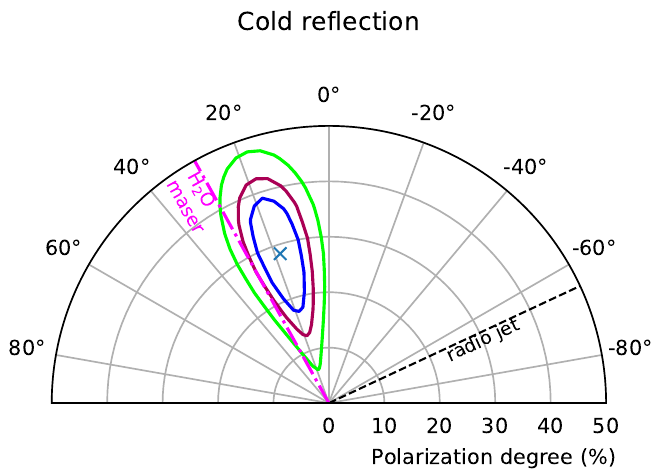}
\caption{Contour plot of the polarization degree and angle for the cold reflector. The blue, red and green lines correspond to the 68, 90 and 99 per cent confidence levels, respectively. The black dashed line marks the direction of the radio jet, while the magenta dash-dotted line marks the direction of the inner H$_2$O maser disc.}
\label{fig:polar}
\end{figure}

We then fit simultaneously the \textit{Chandra} and \textit{IXPE} $I$, $Q$, $U$ Stokes spectra with a model composed of a power law (for the warm reflector continuum), a {\sc pexrav} \citep{pexrav} component (for the scattered cold reflection continuum) and a number of gaussian lines.\footnote{The choice of using the {\sc pexrav} model plus emission lines instead of more sophisticated models which treat continuum and lines together is just due to the fact that the polarization of the continuum reflected emission and of the emission lines are expected to be different.}
We fix the photon index of \textsc{pexrav} at 1.6, consistent with \textit{BeppoSAX}, \textit{Suzaku} and \textit{NuSTAR} previous measurements \citep{matt1999,yang2009,arevalo2014}.
We also fix the energy and flux of the gaussian lines to those derived from the $Chandra/HETG$ spectrum \citep[not reported here for the sake of brevity; see Table 2 of][]{massaro2006}. No strong variability is expected for these lines, since they are associated with extended emission at kpc scales from the central nuclear source.
The warm reflector is partly ionized and it produces a complex emission \citep[e.g.][]{guainazzi1999,kallman2014}.
However, it is phenomenologically well described by a soft power law \citep{bianchi2001,marinucci2013,arevalo2014}, as often found in obscured Seyfert 2 galaxies \citep[e.g.][]{bianchi2001,matt2003,matt2013,bauer2015}. We fix its photon index at 3.0 (see Appendix).
For the \textit{IXPE} spectra, we also include the best-fitting model for the ULXs CG X-1 and CG X-2 derived from \textit{Chandra}, keeping all spectral parameters fixed (see Appendix). The continuum components are multiplied by the {\sc polconst} model, which provides the (energy-independent) polarization degree and angle of each component. We assume the emission lines and the ULXs to be unpolarized.\footnote{If we fit for the polarization of the Fe K$\alpha$ line complex, we obtain an upper limit of 20 per cent, consistent with the expectation for unpolarized fluorescence emission. On the other hand, the Compton shoulder, due to the downscattering of line photons, could theoretically be polarized up to 15 per cent \citep[as computed with the code of][]{ghisellini1994}. However, its flux is at most 20 per cent of that of the line core \citep{bianchi2002,molendi2003}. Therefore, the overall polarization of the line complex should be in any case less than 3 per cent, and we choose to neglect it in the spectral fit.} 
In {\sc xspec}, the model is:
%\begin{align*}
%\textsc{tbabs} \times 
%[  
%\textsc{polconst}^{(0)} 
%\times 
%(
%\sum \textsc{zgauss}^{(i)}
%+ \textsc{powerlaw} \\
%+ \textsc{raymond} 
%+ \textsc{zgauss}
%)
%+ \textsc{polconst}^{(w)} 
%\times 
%\textsc{powerlaw}\\
%+ \textsc{polconst}^{(c)}
%\times
%\textsc{pexrav}
%]
%\end{align*}
\begin{align*}
\textsc{c\_cal} \times
\textsc{tbabs} \times &[  
\textsc{polconst}^{(0)} 
\times 
(
\sum \textsc{zgauss}^{(i)} && \text{lines} \\
&+ \textsc{powerlaw} && \text{CG X-1} \\
& + \textsc{raymond} 
+ \textsc{zgauss}
) && \text{CG X-2} \\
&+ \textsc{polconst}^{(w)} 
\times 
\textsc{powerlaw} && \text{warm refl.}\\
& + \textsc{polconst}^{(c)}
\times
\textsc{pexrav} 
] && \text{cold refl.} 
\end{align*}
where \textsc{c\_cal} denotes the cross-calibration constant. We keep the polarization degree of \textsc{polconst}$^{(0)}$ fixed at zero, which is needed to assume an unpolarized component in \textsc{xspec}. Assuming that the warm and cold reflectors are both axially symmetric structures, their polarization angle is expected to be either parallel or perpendicular to the common symmetry axis. For simplicity, we assume the same polarization angle for the warm and cold reflectors. 
However, we obtain consistent results assuming orthogonal vectors, or by leaving the two polarization angles untied. 
The fit is very good, with $\chi^2$/d.o.f.=702/662. The data and best-fitting model are shown in Figs.~\ref{fig:bestfit} and \ref{fig:model}. The best-fitting parameters, and in particular the polarization degree and angle, are reported in Table~\ref{tab:Fit} and Fig.\,\ref{fig:polar}. 
The polarization degree of the warm reflector is unconstrained, due to its relatively low flux ($20\pm 5$ per cent of the total flux in the 2--4 keV band, see also Fig.\,\ref{fig:model}). For the cold reflector, instead, we obtain significant constraints: a polarization degree of $28\pm 7$ per cent and a polarization angle of $18\degr\pm 5\degr$ (68 per cent confidence for one parameter; see also the two-parameter contours in Fig.\,\ref{fig:polar}).  Interestingly, the polarization angle is consistent with being  perpendicular to the radio jet \citep[position angle of 295\degr,][]{elmouttie1998b}, which also roughly coincides with the axis of the H$\alpha$ ionization cone \citep{elmouttie1998a}, and is close to the direction of the inner H$_2$O maser disc \citep[position angle of $29\degr\pm3\degr$,][]{greenhill2003}.

We also note that the ULXs could in principle be polarized, however they are subdominant at all energies (see Fig. \ref{fig:model}). Even if we assume a polarization degree of 10 per cent for the ULXs, the fit discussed above is essentially unchanged. Thus, unless highly polarized \citep[which is unlikely, see e.g.][]{ixpe_casA,ixpe_cygX-1}, the ULXs cannot affect the AGN measurement. 
%For example, from the \textit{IXPE} data \cite{ixpe_casA} find a polarization degree of $1.8 \pm 0.3$ per cent in the supernova remnant Cassiopeia A, while \cite{ixpe_cygX-1} find $4.0 \pm 0.2$ per cent in the black hole X-ray binary Cygnus X-1. Even if we assume a polarization degree of 10 per cent for the ULXs in the Circinus field, the fit discussed above is essentially unchanged. 

\begin{figure*}
\includegraphics[width=1.0\columnwidth]{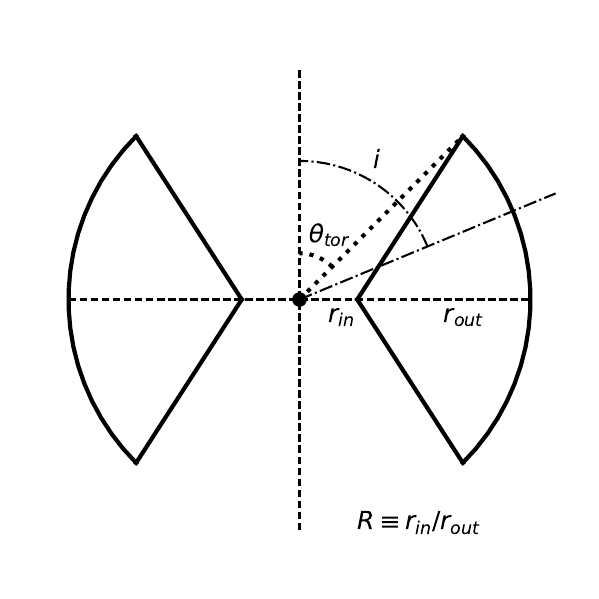}
\includegraphics[width=1.0\columnwidth]{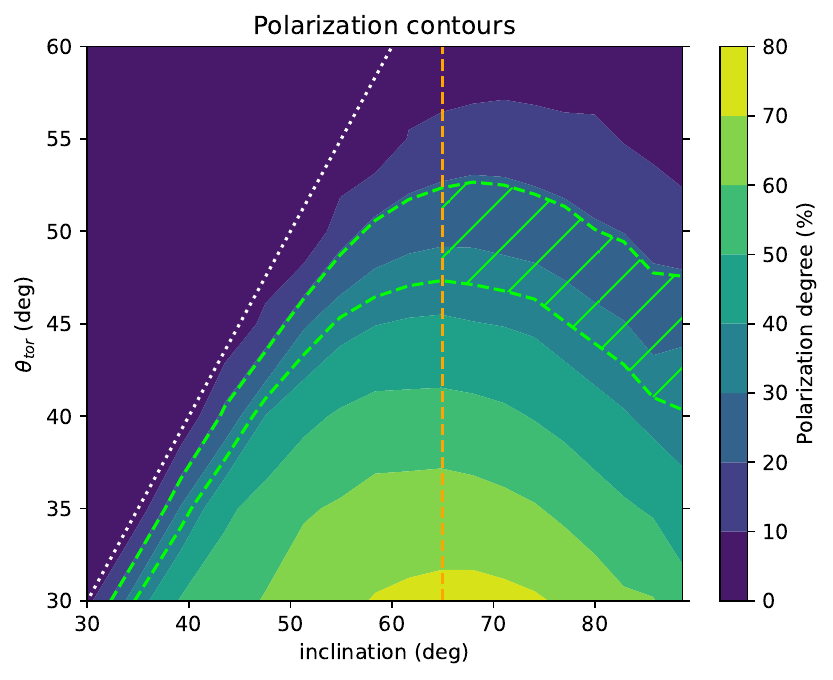}
\caption{\textit{Left panel:}
Sketch of the torus geometry assumed for polarimetric modelling. The torus opening angle $\theta_{\rm tor}$ and the observer inclination $i$ are both measured from the symmetry axis.
\textit{Right panel:}
Contour plots of the constant polarization degree calculated for scattering off a torus, on the $\theta_{\rm tor}$ -- $i$ plane. The white dotted line corresponds to the equality $\theta_{\rm tor} = i$; the condition for the obscuration of the central source is $\theta_{\rm tor} < i$. 
The green dashed curves enclose the 68 per cent confidence level region on the measured polarization for the cold reflector in the Circinus galaxy.
The orange dashed line marks the inclination of the host galaxy, likely a lower limit to the torus inclination (as discussed in the text). The green hatches mark the region of the parameter space consistent with all the observational constraints.
}
\label{fig:simulations}
\end{figure*}

\section{Discussion and conclusions}
\label{sec:disc}

The radiation reflected off the torus, assumed to have an axial symmetry,
is expected to be highly polarized \citep{ghisellini1994,goosmann&matt2011,marin2018_sy2,ratheesh2021}, with a polarization degree depending on the geometrical parameters (namely the torus aperture and the system's inclination) and a polarization angle orthogonal to the torus axis. To better explore the polarization properties of the reflecting torus, we perform calculations with the Monte Carlo radiative transfer code described in \cite{ghisellini1994}. The code takes into account Compton scattering and photoelectric absorption in a neutral medium with solar abundances \citep[see also][]{matt1991}. \cite{ratheesh2021} applied the code to compute the polarization degree of radiation scattered by a torus-like distribution. We show in Fig. \ref{fig:simulations} (left panel) a sketch of the assumed geometry. The shape of the torus is defined by the half-opening angle $\theta_{\rm tor}$ and the ratio $R$ between the inner and outer radius. Like \cite{ratheesh2021}, we assume two different values $R=0.1$ and 0.5. As for the equatorial column density, we assume $N_{\rm H} = 1 \times 10^{25}$\,cm$^{-2}$, consistent with the X-ray measurements of \cite{arevalo2014}.

In Fig. \ref{fig:simulations} (right panel), we show the contours of the polarization degree, calculated as a function of the inclination and of the torus aperture. The 68 per cent confidence level region for the observed value of polarization is also shown. We only plot the case with $R=0.1$, because $R=0.5$ yields very similar results. Assuming an observer inclination of 65\degr, namely that of the host galaxy \citep{freeman1977}, we would obtain a torus aperture roughly in the range 45\degr--55\degr. 
However, the actual inclination might be larger, and different measurements are consistent with an edge-on torus. From the CO(3-2) emission within the central 10\,pc, \cite{izumi2018} estimate $i>70\degr$ and note that the inclination could increase at lower radii. 
Indeed, for the 1-pc scale mid-infrared disc, \cite{tristram2014} estimate $i>75\degr$, while \cite{isbell2022} report $i>83\degr$. As suggested by \cite{izumi2018} and \cite{isbell2022}, the inclination could reach 90\degr\ for the H$_2$O maser disc \citep{greenhill2003}. Assuming a torus close to edge-on, we obtain an aperture of 40\degr--50\degr.
Finally, \cite{elmouttie1998a} estimate a value of 33\degr\ for the half-opening angle of the H$\alpha$ cone, while \cite{fischer2013} estimate 41\degr\ for the [O\,{\sc iii}] cone. This implies a lower limit to the torus aperture which is consistent with our calculations.  

A torus-like distribution of gas and dust surrounding the central engine is the cornerstone of AGN Unification Models, notwithstanding their continuing development \citep[e.g.][]{antonucci&miller1985,antonucci1993,bianchi2012,marin2014,netzer2015,marin2016,ramos-almeida&ricci2017}. In the current understanding, the torus is actually a clumpy structure, as was early on suggested by \cite{krolik&begelman1988}. Especially in the case of the Circinus galaxy, this is strongly supported by infrared interferometry \citep[][]{tristram2007,tristram2014,stalevski2017,stalevski2019,isbell2022} and by the imaging analysis of the X-ray reflector (\citealt{marinucci2013}; see also \citealt{andonie2022}). 
In any case, the X-ray polarimetric properties are in qualitative agreement with the expected results for a geometrically thick torus having an aspect ratio $h/r \sim 1$, meaning a covering factor $\cos \theta_{\rm tor} \sim 0.7$. This result is consistent with the estimate obtained for local AGNs from the average X-ray absorption properties \citep[see][and references therein]{ramos-almeida&ricci2017}.
%\red{AI:How does this compare with population statistics of type 1 vs 2 AGN?}. 

We note that the Unification Model was introduced after the detection of broad emission lines in the optical polarized spectrum of Seyfert 2 galaxies, with the first one being the Compton-thick AGN NGC~1068 (\citealt{antonucci&miller1985}; see also \citealt{oliva1998} for the case of the Circinus galaxy). %X-ray polarimetry now confirms the basic prediction of this scenario, and paves the way for future multiwavelength polarimetric studies of AGNs \citep[e.g.][]{marin2018_ngc1068}.
%For example, the polarization properties of the warm reflection component remain an open problem. 
%Moreover, modeling the total and polarized spectral energy distribution from the infrared to the X-ray band will likely yield strong constraints on the geometry of the system.
%Further observations of Compton-thick sources will allow us to study the AGN circumnuclear regions with increasing detail.
Adding a crucial piece of information, X-ray polarimetry now confirms the basic picture of this scenario.
A comprehensive discussion of its implications, taking into account polarimetric information at all wavelengths \citep[e.g.][]{marin2018_ngc1068}, is beyond the scope of this paper and will be the subject of a future work.

%Polarization angle and position of radio jet and ionization cone

\section*{Acknowledgements}

The Imaging X ray Polarimetry Explorer (IXPE) is a joint US and Italian mission.  The US contribution is supported by the National Aeronautics and Space Administration (NASA) and led and managed by its Marshall Space Flight Center (MSFC), with industry partner Ball Aerospace (contract NNM15AA18C).  The Italian contribution is supported by the Italian Space Agency (Agenzia Spaziale Italiana, ASI) through contract ASI-OHBI-2017-12-I.0, agreements ASI-INAF-2017-12-H0 and ASI-INFN-2017.13-H0, and its Space Science Data Center (SSDC) with agreements ASI-INAF-2022-14-HH.0 and ASI-INFN 2021-43-HH.0, and by the Istituto Nazionale di Astrofisica (INAF) and the Istituto Nazionale di Fisica Nucleare (INFN) in Italy.  This research used data products provided by the IXPE Team (MSFC, SSDC, INAF, and INFN) and distributed with additional software tools by the High-Energy Astrophysics Science Archive Research Center (HEASARC), at NASA Goddard Space Flight Center (GSFC).
JPou and SST were supported by the Academy of Finland grants   349373 and 349906.
AIng acknowledges support from the Royal Society.
MDov, JSvo and VKar acknowledge the support from the GACR project 21-06825X.
POP acknowledges financial support from the French High Energy Program (PNHE/CNRS) and from the french spatial agency (CNES).
%%%%%%%%%%%%%%%%%%%%%%%%%%%%%%%%%%%%%%%%%%%%%%%%%%
\section*{Data Availability}
The data used in this paper are publicly available in the HEASARC database. The code by \cite{ghisellini1994} used to perform the numerical simulations is proprietary; simulation data supporting the findings of the article will be shared on reasonable request.

\bibliographystyle{mnras}
\bibliography{bibliography_circinus} 

\begin{thebibliography}{}
\makeatletter
\relax
\def\mn@urlcharsother{\let\do\@makeother \do\$\do\&\do\#\do\^\do\_\do\%\do\~}
\def\mn@doi{\begingroup\mn@urlcharsother \@ifnextchar [ {\mn@doi@}
  {\mn@doi@[]}}
\def\mn@doi@[#1]#2{\def\@tempa{#1}\ifx\@tempa\@empty \href
  {http://dx.doi.org/#2} {doi:#2}\else \href {http://dx.doi.org/#2} {#1}\fi
  \endgroup}
\def\mn@eprint#1#2{\mn@eprint@#1:#2::\@nil}
\def\mn@eprint@arXiv#1{\href {http://arxiv.org/abs/#1} {{\tt arXiv:#1}}}
\def\mn@eprint@dblp#1{\href {http://dblp.uni-trier.de/rec/bibtex/#1.xml}
  {dblp:#1}}
\def\mn@eprint@#1:#2:#3:#4\@nil{\def\@tempa {#1}\def\@tempb {#2}\def\@tempc
  {#3}\ifx \@tempc \@empty \let \@tempc \@tempb \let \@tempb \@tempa \fi \ifx
  \@tempb \@empty \def\@tempb {arXiv}\fi \@ifundefined
  {mn@eprint@\@tempb}{\@tempb:\@tempc}{\expandafter \expandafter \csname
  mn@eprint@\@tempb\endcsname \expandafter{\@tempc}}}

\bibitem[\protect\citeauthoryear{{Andonie}, {Ricci}, {Paltani}, {Ar{\'e}valo},
  {Treister}, {Bauer}  \& {Stalevski}}{{Andonie} et~al.}{2022}]{andonie2022}
{Andonie} C.,  {Ricci} C.,  {Paltani} S.,  {Ar{\'e}valo} P.,  {Treister} E.,
  {Bauer} F.,   {Stalevski} M.,  2022, \mn@doi [\mnras]
  {10.1093/mnras/stac403}, \href
  {https://ui.adsabs.harvard.edu/abs/2022MNRAS.511.5768A} {511, 5768}

\bibitem[\protect\citeauthoryear{{Antonucci}}{{Antonucci}}{1993}]{antonucci1993}
{Antonucci} R.,  1993, \mn@doi [\araa] {10.1146/annurev.aa.31.090193.002353},
  \href {https://ui.adsabs.harvard.edu/abs/1993ARA&A..31..473A} {31, 473}

\bibitem[\protect\citeauthoryear{{Antonucci} \& {Miller}}{{Antonucci} \&
  {Miller}}{1985}]{antonucci&miller1985}
{Antonucci} R.~R.~J.,  {Miller} J.~S.,  1985, \mn@doi [\apj] {10.1086/163559},
  \href {https://ui.adsabs.harvard.edu/abs/1985ApJ...297..621A} {297, 621}

\bibitem[\protect\citeauthoryear{{Ar{\'e}valo} et~al.,}{{Ar{\'e}valo}
  et~al.}{2014}]{arevalo2014}
{Ar{\'e}valo} P.,  et~al., 2014, \mn@doi [\apj] {10.1088/0004-637X/791/2/81},
  \href {https://ui.adsabs.harvard.edu/abs/2014ApJ...791...81A} {791, 81}

\bibitem[\protect\citeauthoryear{{Arnaud}}{{Arnaud}}{1996}]{xspec}
{Arnaud} K.~A.,  1996, in {Jacoby} G.~H.,  {Barnes} J.,  eds,  Astronomical
  Society of the Pacific Conference Series Vol. 101, Astronomical Data Analysis
  Software and Systems V. p.~17

\bibitem[\protect\citeauthoryear{{Baldini}, {Bucciantini}, {Di Lalla},
  {Ehlert}, {Manfreda}, {Omodei}, {Pesce-Rollins}  \& {Sgr{\`o}}}{{Baldini}
  et~al.}{2022}]{baldini22}
{Baldini} L.,  {Bucciantini} N.,  {Di Lalla} N.,  {Ehlert} S.~R.,  {Manfreda}
  A.,  {Omodei} N.,  {Pesce-Rollins} M.,   {Sgr{\`o}} C.,  2022, arXiv
  e-prints, \href {https://ui.adsabs.harvard.edu/abs/2022arXiv220306384B} {p.
  arXiv:2203.06384}

\bibitem[\protect\citeauthoryear{{Bauer}, {Brandt}, {Sambruna}, {Chartas},
  {Garmire}, {Kaspi}  \& {Netzer}}{{Bauer} et~al.}{2001}]{bauer2001}
{Bauer} F.~E.,  {Brandt} W.~N.,  {Sambruna} R.~M.,  {Chartas} G.,  {Garmire}
  G.~P.,  {Kaspi} S.,   {Netzer} H.,  2001, \mn@doi [\aj] {10.1086/321123},
  \href {https://ui.adsabs.harvard.edu/abs/2001AJ....122..182B} {122, 182}

\bibitem[\protect\citeauthoryear{{Bauer} et~al.,}{{Bauer}
  et~al.}{2015}]{bauer2015}
{Bauer} F.~E.,  et~al., 2015, \mn@doi [\apj] {10.1088/0004-637X/812/2/116},
  \href {https://ui.adsabs.harvard.edu/abs/2015ApJ...812..116B} {812, 116}

\bibitem[\protect\citeauthoryear{{Bianchi}, {Matt}  \& {Iwasawa}}{{Bianchi}
  et~al.}{2001}]{bianchi2001}
{Bianchi} S.,  {Matt} G.,   {Iwasawa} K.,  2001, \mn@doi [\mnras]
  {10.1046/j.1365-8711.2001.04156.x}, \href
  {https://ui.adsabs.harvard.edu/abs/2001MNRAS.322..669B} {322, 669}

\bibitem[\protect\citeauthoryear{{Bianchi}, {Matt}, {Fiore}, {Fabian},
  {Iwasawa}  \& {Nicastro}}{{Bianchi} et~al.}{2002}]{bianchi2002}
{Bianchi} S.,  {Matt} G.,  {Fiore} F.,  {Fabian} A.~C.,  {Iwasawa} K.,
  {Nicastro} F.,  2002, \mn@doi [\aap] {10.1051/0004-6361:20021414}, \href
  {https://ui.adsabs.harvard.edu/abs/2002A&A...396..793B} {396, 793}

\bibitem[\protect\citeauthoryear{{Bianchi}, {Maiolino}  \&
  {Risaliti}}{{Bianchi} et~al.}{2012}]{bianchi2012}
{Bianchi} S.,  {Maiolino} R.,   {Risaliti} G.,  2012, \mn@doi [Advances in
  Astronomy] {10.1155/2012/782030}, \href
  {https://ui.adsabs.harvard.edu/abs/2012AdAst2012E..17B} {2012, 782030}

\bibitem[\protect\citeauthoryear{{Brinkmann}, {Siebert}  \&
  {Boller}}{{Brinkmann} et~al.}{1994}]{brinkmann1994}
{Brinkmann} W.,  {Siebert} J.,   {Boller} T.,  1994, \aap, \href
  {https://ui.adsabs.harvard.edu/abs/1994A&A...281..355B} {281, 355}

\bibitem[\protect\citeauthoryear{{Curran}, {Koribalski}  \& {Bains}}{{Curran}
  et~al.}{2008}]{curran2008}
{Curran} S.~J.,  {Koribalski} B.~S.,   {Bains} I.,  2008, \mn@doi [\mnras]
  {10.1111/j.1365-2966.2008.13574.x}, \href
  {https://ui.adsabs.harvard.edu/abs/2008MNRAS.389...63C} {389, 63}

\bibitem[\protect\citeauthoryear{{Di Marco} et~al.,}{{Di Marco}
  et~al.}{2022}]{dimarco22}
{Di Marco} A.,  et~al., 2022, \mn@doi [\aj] {10.3847/1538-3881/ac51c9}, \href
  {https://ui.adsabs.harvard.edu/abs/2022AJ....163..170D} {163, 170}

\bibitem[\protect\citeauthoryear{{Elmouttie}, {Koribalski}, {Gordon}, {Taylor},
  {Houghton}, {Lavezzi}, {Haynes}  \& {Jones}}{{Elmouttie}
  et~al.}{1998a}]{elmouttie1998a}
{Elmouttie} M.,  {Koribalski} B.,  {Gordon} S.,  {Taylor} K.,  {Houghton} S.,
  {Lavezzi} T.,  {Haynes} R.,   {Jones} K.,  1998a, \mn@doi [\mnras]
  {10.1046/j.1365-8711.1998.01402.x}, \href
  {https://ui.adsabs.harvard.edu/abs/1998MNRAS.297...49E} {297, 49}

\bibitem[\protect\citeauthoryear{{Elmouttie}, {Haynes}, {Jones}, {Sadler}  \&
  {Ehle}}{{Elmouttie} et~al.}{1998b}]{elmouttie1998b}
{Elmouttie} M.,  {Haynes} R.~F.,  {Jones} K.~L.,  {Sadler} E.~M.,   {Ehle} M.,
  1998b, \mn@doi [\mnras] {10.1046/j.1365-8711.1998.01592.x}, \href
  {https://ui.adsabs.harvard.edu/abs/1998MNRAS.297.1202E} {297, 1202}

\bibitem[\protect\citeauthoryear{{Esposito}, {Israel}, {Milisavljevic},
  {Mapelli}, {Zampieri}, {Sidoli}, {Fabbiano}  \& {Rodr{\'\i}guez
  Castillo}}{{Esposito} et~al.}{2015}]{esposito2015}
{Esposito} P.,  {Israel} G.~L.,  {Milisavljevic} D.,  {Mapelli} M.,  {Zampieri}
  L.,  {Sidoli} L.,  {Fabbiano} G.,   {Rodr{\'\i}guez Castillo} G.~A.,  2015,
  \mn@doi [\mnras] {10.1093/mnras/stv1379}, \href
  {https://ui.adsabs.harvard.edu/abs/2015MNRAS.452.1112E} {452, 1112}

\bibitem[\protect\citeauthoryear{{Fischer}, {Crenshaw}, {Kraemer}  \&
  {Schmitt}}{{Fischer} et~al.}{2013}]{fischer2013}
{Fischer} T.~C.,  {Crenshaw} D.~M.,  {Kraemer} S.~B.,   {Schmitt} H.~R.,  2013,
  \mn@doi [\apjs] {10.1088/0067-0049/209/1/1}, \href
  {https://ui.adsabs.harvard.edu/abs/2013ApJS..209....1F} {209, 1}

\bibitem[\protect\citeauthoryear{{Freeman}, {Karlsson}, {Lynga}, {Burrell},
  {van Woerden}, {Goss}  \& {Mebold}}{{Freeman} et~al.}{1977}]{freeman1977}
{Freeman} K.~C.,  {Karlsson} B.,  {Lynga} G.,  {Burrell} J.~F.,  {van Woerden}
  H.,  {Goss} W.~M.,   {Mebold} U.,  1977, \aap, \href
  {https://ui.adsabs.harvard.edu/abs/1977A&A....55..445F} {55, 445}

\bibitem[\protect\citeauthoryear{{Fruscione} et~al.,}{{Fruscione}
  et~al.}{2006}]{ciao}
{Fruscione} A.,  et~al., 2006, in {Silva} D.~R.,  E. D.~R.,  eds,  SPIE Conf.
  Ser. Vol. 6270, Observatory Operations: Strategies, Processes, and Systems.
  p. 62701V, \mn@doi{10.1117/12.671760}

\bibitem[\protect\citeauthoryear{{Garmire}, {Bautz}, {Ford}, {Nousek}  \&
  {Ricker}}{{Garmire} et~al.}{2003}]{acis}
{Garmire} G.~P.,  {Bautz} M.~W.,  {Ford} P.~G.,  {Nousek} J.~A.,   {Ricker}
  G.~R.,  2003, in {Truemper} J.~E.,  D. T.~H.,  eds,  SPIE Conf. Ser. Vol.
  4851, X-Ray and Gamma-Ray Telescopes and Instruments for Astronomy. pp 28--44

\bibitem[\protect\citeauthoryear{{Ghisellini}, {Haardt}  \&
  {Matt}}{{Ghisellini} et~al.}{1994}]{ghisellini1994}
{Ghisellini} G.,  {Haardt} F.,   {Matt} G.,  1994, \mn@doi [\mnras]
  {10.1093/mnras/267.3.743}, \href
  {https://ui.adsabs.harvard.edu/abs/1994MNRAS.267..743G} {267, 743}

\bibitem[\protect\citeauthoryear{{Goosmann} \& {Matt}}{{Goosmann} \&
  {Matt}}{2011}]{goosmann&matt2011}
{Goosmann} R.~W.,  {Matt} G.,  2011, \mn@doi [\mnras]
  {10.1111/j.1365-2966.2011.18923.x}, \href
  {https://ui.adsabs.harvard.edu/abs/2011MNRAS.415.3119G} {415, 3119}

\bibitem[\protect\citeauthoryear{{Greenhill} et~al.,}{{Greenhill}
  et~al.}{2003}]{greenhill2003}
{Greenhill} L.~J.,  et~al., 2003, \mn@doi [\apj] {10.1086/374862}, \href
  {https://ui.adsabs.harvard.edu/abs/2003ApJ...590..162G} {590, 162}

\bibitem[\protect\citeauthoryear{{Guainazzi} et~al.,}{{Guainazzi}
  et~al.}{1999}]{guainazzi1999}
{Guainazzi} M.,  et~al., 1999, \mn@doi [\mnras]
  {10.1046/j.1365-8711.1999.02803.x}, \href
  {https://ui.adsabs.harvard.edu/abs/1999MNRAS.310...10G} {310, 10}

\bibitem[\protect\citeauthoryear{{Isbell} et~al.,}{{Isbell}
  et~al.}{2022}]{isbell2022}
{Isbell} J.~W.,  et~al., 2022, \mn@doi [\aap] {10.1051/0004-6361/202243271},
  \href {https://ui.adsabs.harvard.edu/abs/2022A&A...663A..35I} {663, A35}

\bibitem[\protect\citeauthoryear{{Izumi}, {Wada}, {Fukushige}, {Hamamura}  \&
  {Kohno}}{{Izumi} et~al.}{2018}]{izumi2018}
{Izumi} T.,  {Wada} K.,  {Fukushige} R.,  {Hamamura} S.,   {Kohno} K.,  2018,
  \mn@doi [\apj] {10.3847/1538-4357/aae20b}, \href
  {https://ui.adsabs.harvard.edu/abs/2018ApJ...867...48I} {867, 48}

\bibitem[\protect\citeauthoryear{{Kallman}, {Evans}, {Marshall}, {Canizares},
  {Longinotti}, {Nowak}  \& {Schulz}}{{Kallman} et~al.}{2014}]{kallman2014}
{Kallman} T.,  {Evans} D.~A.,  {Marshall} H.,  {Canizares} C.,  {Longinotti}
  A.,  {Nowak} M.,   {Schulz} N.,  2014, \mn@doi [\apj]
  {10.1088/0004-637X/780/2/121}, \href
  {https://ui.adsabs.harvard.edu/abs/2014ApJ...780..121K} {780, 121}

\bibitem[\protect\citeauthoryear{{Kawamuro}, {Izumi}  \& {Imanishi}}{{Kawamuro}
  et~al.}{2019}]{kawamuro2019}
{Kawamuro} T.,  {Izumi} T.,   {Imanishi} M.,  2019, \mn@doi [\pasj]
  {10.1093/pasj/psz045}, \href
  {https://ui.adsabs.harvard.edu/abs/2019PASJ...71...68K} {71, 68}

\bibitem[\protect\citeauthoryear{{Kislat}, {Clark}, {Beilicke}  \&
  {Krawczynski}}{{Kislat} et~al.}{2015}]{kislat2015}
{Kislat} F.,  {Clark} B.,  {Beilicke} M.,   {Krawczynski} H.,  2015, \mn@doi
  [Astroparticle Physics] {10.1016/j.astropartphys.2015.02.007}, \href
  {https://ui.adsabs.harvard.edu/abs/2015APh....68...45K} {68, 45}

\bibitem[\protect\citeauthoryear{{Krawczynski} et~al.,}{{Krawczynski}
  et~al.}{2022}]{ixpe_cygX-1}
{Krawczynski} H.,  et~al., 2022, arXiv e-prints, \href
  {https://ui.adsabs.harvard.edu/abs/2022arXiv220609972K} {p. arXiv:2206.09972}

\bibitem[\protect\citeauthoryear{{Krolik} \& {Begelman}}{{Krolik} \&
  {Begelman}}{1988}]{krolik&begelman1988}
{Krolik} J.~H.,  {Begelman} M.~C.,  1988, \mn@doi [\apj] {10.1086/166414},
  \href {https://ui.adsabs.harvard.edu/abs/1988ApJ...329..702K} {329, 702}

\bibitem[\protect\citeauthoryear{{Magdziarz} \& {Zdziarski}}{{Magdziarz} \&
  {Zdziarski}}{1995}]{pexrav}
{Magdziarz} P.,  {Zdziarski} A.~A.,  1995, \mn@doi [\mnras]
  {10.1093/mnras/273.3.837}, \href
  {https://ui.adsabs.harvard.edu/abs/1995MNRAS.273..837M} {273, 837}

\bibitem[\protect\citeauthoryear{{Malizia}, {Stephen}, {Bassani}, {Bird},
  {Panessa}  \& {Ubertini}}{{Malizia} et~al.}{2009}]{malizia2009}
{Malizia} A.,  {Stephen} J.~B.,  {Bassani} L.,  {Bird} A.~J.,  {Panessa} F.,
  {Ubertini} P.,  2009, \mn@doi [\mnras] {10.1111/j.1365-2966.2009.15330.x},
  \href {https://ui.adsabs.harvard.edu/abs/2009MNRAS.399..944M} {399, 944}

\bibitem[\protect\citeauthoryear{{Marconi}, {Moorwood}, {Origlia}  \&
  {Oliva}}{{Marconi} et~al.}{1994}]{marconi1994}
{Marconi} A.,  {Moorwood} A.~F.~M.,  {Origlia} L.,   {Oliva} E.,  1994, The
  Messenger, \href {https://ui.adsabs.harvard.edu/abs/1994Msngr..78...20M} {78,
  20}

\bibitem[\protect\citeauthoryear{{Marin}}{{Marin}}{2014}]{marin2014}
{Marin} F.,  2014, \mn@doi [\mnras] {10.1093/mnras/stu593}, \href
  {https://ui.adsabs.harvard.edu/abs/2014MNRAS.441..551M} {441, 551}

\bibitem[\protect\citeauthoryear{{Marin}}{{Marin}}{2016}]{marin2016}
{Marin} F.,  2016, \mn@doi [\mnras] {10.1093/mnras/stw1131}, \href
  {https://ui.adsabs.harvard.edu/abs/2016MNRAS.460.3679M} {460, 3679}

\bibitem[\protect\citeauthoryear{{Marin}}{{Marin}}{2018}]{marin2018_ngc1068}
{Marin} F.,  2018, \mn@doi [\mnras] {10.1093/mnras/sty1566}, \href
  {https://ui.adsabs.harvard.edu/abs/2018MNRAS.479.3142M} {479, 3142}

\bibitem[\protect\citeauthoryear{{Marin}, {Dov{\v{c}}iak}, {Muleri}, {Kislat}
  \& {Krawczynski}}{{Marin} et~al.}{2018}]{marin2018_sy2}
{Marin} F.,  {Dov{\v{c}}iak} M.,  {Muleri} F.,  {Kislat} F.~F.,   {Krawczynski}
  H.~S.,  2018, \mn@doi [\mnras] {10.1093/mnras/stx2382}, \href
  {https://ui.adsabs.harvard.edu/abs/2018MNRAS.473.1286M} {473, 1286}

\bibitem[\protect\citeauthoryear{{Marinucci}, {Miniutti}, {Bianchi}, {Matt}  \&
  {Risaliti}}{{Marinucci} et~al.}{2013}]{marinucci2013}
{Marinucci} A.,  {Miniutti} G.,  {Bianchi} S.,  {Matt} G.,   {Risaliti} G.,
  2013, \mn@doi [\mnras] {10.1093/mnras/stt1759}, \href
  {https://ui.adsabs.harvard.edu/abs/2013MNRAS.436.2500M} {436, 2500}

\bibitem[\protect\citeauthoryear{{Marinucci} et~al.,}{{Marinucci}
  et~al.}{2016}]{marinucci2016}
{Marinucci} A.,  et~al., 2016, \mn@doi [\mnras] {10.1093/mnrasl/slv178}, \href
  {https://ui.adsabs.harvard.edu/abs/2016MNRAS.456L..94M} {456, L94}

\bibitem[\protect\citeauthoryear{{Marinucci} et~al.,}{{Marinucci}
  et~al.}{2022}]{marinucci2022}
{Marinucci} A.,  et~al., 2022, \mn@doi [\mnras] {10.1093/mnras/stac2634}, \href
  {https://ui.adsabs.harvard.edu/abs/2022MNRAS.516.5907M} {516, 5907}

\bibitem[\protect\citeauthoryear{{Massaro}, {Bianchi}, {Matt}, {D'Onofrio}  \&
  {Nicastro}}{{Massaro} et~al.}{2006}]{massaro2006}
{Massaro} F.,  {Bianchi} S.,  {Matt} G.,  {D'Onofrio} E.,   {Nicastro} F.,
  2006, \mn@doi [\aap] {10.1051/0004-6361:20054772}, \href
  {https://ui.adsabs.harvard.edu/abs/2006A&A...455..153M} {455, 153}

\bibitem[\protect\citeauthoryear{{Matt}, {Perola}  \& {Piro}}{{Matt}
  et~al.}{1991}]{matt1991}
{Matt} G.,  {Perola} G.~C.,   {Piro} L.,  1991, \aap, \href
  {https://ui.adsabs.harvard.edu/abs/1991A&A...247...25M} {247, 25}

\bibitem[\protect\citeauthoryear{{Matt} et~al.,}{{Matt}
  et~al.}{1996}]{matt1996}
{Matt} G.,  et~al., 1996, \mn@doi [\mnras] {10.1093/mnras/281.4.L69}, \href
  {https://ui.adsabs.harvard.edu/abs/1996MNRAS.281L..69M} {281, L69}

\bibitem[\protect\citeauthoryear{{Matt} et~al.,}{{Matt}
  et~al.}{1999}]{matt1999}
{Matt} G.,  et~al., 1999, \aap, \href
  {https://ui.adsabs.harvard.edu/abs/1999A&A...341L..39M} {341, L39}

\bibitem[\protect\citeauthoryear{{Matt}, {Fabian}, {Guainazzi}, {Iwasawa},
  {Bassani}  \& {Malaguti}}{{Matt} et~al.}{2000}]{matt2000}
{Matt} G.,  {Fabian} A.~C.,  {Guainazzi} M.,  {Iwasawa} K.,  {Bassani} L.,
  {Malaguti} G.,  2000, \mn@doi [\mnras] {10.1046/j.1365-8711.2000.03721.x},
  \href {https://ui.adsabs.harvard.edu/abs/2000MNRAS.318..173M} {318, 173}

\bibitem[\protect\citeauthoryear{{Matt}, {Bianchi}, {Guainazzi}, {Brandt},
  {Fabian}, {Iwasawa}  \& {Perola}}{{Matt} et~al.}{2003}]{matt2003}
{Matt} G.,  {Bianchi} S.,  {Guainazzi} M.,  {Brandt} W.~N.,  {Fabian} A.~C.,
  {Iwasawa} K.,   {Perola} G.~C.,  2003, \mn@doi [\aap]
  {10.1051/0004-6361:20021817}, \href
  {https://ui.adsabs.harvard.edu/abs/2003A&A...399..519M} {399, 519}

\bibitem[\protect\citeauthoryear{{Matt}, {Bianchi}, {Marinucci}, {Guainazzi},
  {Iwawasa}  \& {Jimenez Bailon}}{{Matt} et~al.}{2013}]{matt2013}
{Matt} G.,  {Bianchi} S.,  {Marinucci} A.,  {Guainazzi} M.,  {Iwawasa} K.,
  {Jimenez Bailon} E.,  2013, \mn@doi [\aap] {10.1051/0004-6361/201321293},
  \href {https://ui.adsabs.harvard.edu/abs/2013A&A...556A..91M} {556, A91}

\bibitem[\protect\citeauthoryear{{Molendi}, {Bianchi}  \& {Matt}}{{Molendi}
  et~al.}{2003}]{molendi2003}
{Molendi} S.,  {Bianchi} S.,   {Matt} G.,  2003, \mn@doi [\mnras]
  {10.1046/j.1365-8711.2003.06783.x}, \href
  {https://ui.adsabs.harvard.edu/abs/2003MNRAS.343L...1M} {343, L1}

\bibitem[\protect\citeauthoryear{{Netzer}}{{Netzer}}{2015}]{netzer2015}
{Netzer} H.,  2015, \mn@doi [\araa] {10.1146/annurev-astro-082214-122302},
  \href {https://ui.adsabs.harvard.edu/abs/2015ARA&A..53..365N} {53, 365}

\bibitem[\protect\citeauthoryear{{Oliva}, {Marconi}, {Cimatti}  \& {di Serego
  Alighieri}}{{Oliva} et~al.}{1998}]{oliva1998}
{Oliva} E.,  {Marconi} A.,  {Cimatti} A.,   {di Serego Alighieri} S.,  1998,
  \aap, \href {https://ui.adsabs.harvard.edu/abs/1998A&A...329L..21O} {329,
  L21}

\bibitem[\protect\citeauthoryear{{Piconcelli}, {Jimenez-Bail{\'o}n},
  {Guainazzi}, {Schartel}, {Rodr{\'\i}guez-Pascual}  \&
  {Santos-Lle{\'o}}}{{Piconcelli} et~al.}{2004}]{piconcelli2004}
{Piconcelli} E.,  {Jimenez-Bail{\'o}n} E.,  {Guainazzi} M.,  {Schartel} N.,
  {Rodr{\'\i}guez-Pascual} P.~M.,   {Santos-Lle{\'o}} M.,  2004, \mn@doi
  [\mnras] {10.1111/j.1365-2966.2004.07764.x}, \href
  {https://ui.adsabs.harvard.edu/abs/2004MNRAS.351..161P} {351, 161}

\bibitem[\protect\citeauthoryear{{Qiu} et~al.,}{{Qiu} et~al.}{2019}]{qiu2019}
{Qiu} Y.,  et~al., 2019, \mn@doi [\apj] {10.3847/1538-4357/ab16e7}, \href
  {https://ui.adsabs.harvard.edu/abs/2019ApJ...877...57Q} {877, 57}

\bibitem[\protect\citeauthoryear{{Ramos Almeida} \& {Ricci}}{{Ramos Almeida} \&
  {Ricci}}{2017}]{ramos-almeida&ricci2017}
{Ramos Almeida} C.,  {Ricci} C.,  2017, \mn@doi [Nature Astronomy]
  {10.1038/s41550-017-0232-z}, \href
  {https://ui.adsabs.harvard.edu/abs/2017NatAs...1..679R} {1, 679}

\bibitem[\protect\citeauthoryear{{Ratheesh}, {Matt}, {Tombesi}, {Soffitta},
  {Pesce-Rollins}  \& {Di Marco}}{{Ratheesh} et~al.}{2021}]{ratheesh2021}
{Ratheesh} A.,  {Matt} G.,  {Tombesi} F.,  {Soffitta} P.,  {Pesce-Rollins} M.,
   {Di Marco} A.,  2021, \mn@doi [\aap] {10.1051/0004-6361/202140701}, \href
  {https://ui.adsabs.harvard.edu/abs/2021A&A...655A..96R} {655, A96}

\bibitem[\protect\citeauthoryear{{Ricci}, {Ueda}, {Koss}, {Trakhtenbrot},
  {Bauer}  \& {Gandhi}}{{Ricci} et~al.}{2015}]{ricci2015}
{Ricci} C.,  {Ueda} Y.,  {Koss} M.~J.,  {Trakhtenbrot} B.,  {Bauer} F.~E.,
  {Gandhi} P.,  2015, \mn@doi [\apjl] {10.1088/2041-8205/815/1/L13}, \href
  {https://ui.adsabs.harvard.edu/abs/2015ApJ...815L..13R} {815, L13}

\bibitem[\protect\citeauthoryear{{Sambruna}, {Brandt}, {Chartas}, {Netzer},
  {Kaspi}, {Garmire}, {Nousek}  \& {Weaver}}{{Sambruna}
  et~al.}{2001a}]{sambruna2001_im}
{Sambruna} R.~M.,  {Brandt} W.~N.,  {Chartas} G.,  {Netzer} H.,  {Kaspi} S.,
  {Garmire} G.~P.,  {Nousek} J.~A.,   {Weaver} K.~A.,  2001a, \mn@doi [\apjl]
  {10.1086/318067}, \href
  {https://ui.adsabs.harvard.edu/abs/2001ApJ...546L...9S} {546, L9}

\bibitem[\protect\citeauthoryear{{Sambruna}, {Netzer}, {Kaspi}, {Brandt},
  {Chartas}, {Garmire}, {Nousek}  \& {Weaver}}{{Sambruna}
  et~al.}{2001b}]{sambruna2001_hrsp}
{Sambruna} R.~M.,  {Netzer} H.,  {Kaspi} S.,  {Brandt} W.~N.,  {Chartas} G.,
  {Garmire} G.~P.,  {Nousek} J.~A.,   {Weaver} K.~A.,  2001b, \mn@doi [\apjl]
  {10.1086/318068}, \href
  {https://ui.adsabs.harvard.edu/abs/2001ApJ...546L..13S} {546, L13}

\bibitem[\protect\citeauthoryear{{Smith} \& {Wilson}}{{Smith} \&
  {Wilson}}{2001}]{smith&wilson2001}
{Smith} D.~A.,  {Wilson} A.~S.,  2001, \mn@doi [\apj] {10.1086/321667}, \href
  {https://ui.adsabs.harvard.edu/abs/2001ApJ...557..180S} {557, 180}

\bibitem[\protect\citeauthoryear{{Soffitta} et~al.,}{{Soffitta}
  et~al.}{2021}]{soffitta2021}
{Soffitta} P.,  et~al., 2021, \mn@doi [\aj] {10.3847/1538-3881/ac19b0}, \href
  {https://ui.adsabs.harvard.edu/abs/2021AJ....162..208S} {162, 208}

\bibitem[\protect\citeauthoryear{{Stalevski}, {Asmus}  \&
  {Tristram}}{{Stalevski} et~al.}{2017}]{stalevski2017}
{Stalevski} M.,  {Asmus} D.,   {Tristram} K. R.~W.,  2017, \mn@doi [\mnras]
  {10.1093/mnras/stx2227}, \href
  {https://ui.adsabs.harvard.edu/abs/2017MNRAS.472.3854S} {472, 3854}

\bibitem[\protect\citeauthoryear{{Stalevski}, {Tristram}  \&
  {Asmus}}{{Stalevski} et~al.}{2019}]{stalevski2019}
{Stalevski} M.,  {Tristram} K. R.~W.,   {Asmus} D.,  2019, \mn@doi [\mnras]
  {10.1093/mnras/stz220}, \href
  {https://ui.adsabs.harvard.edu/abs/2019MNRAS.484.3334S} {484, 3334}

\bibitem[\protect\citeauthoryear{{Torres-Alb{\`a}} et~al.,}{{Torres-Alb{\`a}}
  et~al.}{2021}]{torres-alba2021}
{Torres-Alb{\`a}} N.,  et~al., 2021, \mn@doi [\apj] {10.3847/1538-4357/ac1c73},
  \href {https://ui.adsabs.harvard.edu/abs/2021ApJ...922..252T} {922, 252}

\bibitem[\protect\citeauthoryear{{Tristram} et~al.,}{{Tristram}
  et~al.}{2007}]{tristram2007}
{Tristram} K.~R.~W.,  et~al., 2007, \mn@doi [\aap]
  {10.1051/0004-6361:20078369}, \href
  {https://ui.adsabs.harvard.edu/abs/2007A&A...474..837T} {474, 837}

\bibitem[\protect\citeauthoryear{{Tristram}, {Burtscher}, {Jaffe},
  {Meisenheimer}, {H{\"o}nig}, {Kishimoto}, {Schartmann}  \&
  {Weigelt}}{{Tristram} et~al.}{2014}]{tristram2014}
{Tristram} K. R.~W.,  {Burtscher} L.,  {Jaffe} W.,  {Meisenheimer} K.,
  {H{\"o}nig} S.~F.,  {Kishimoto} M.,  {Schartmann} M.,   {Weigelt} G.,  2014,
  \mn@doi [\aap] {10.1051/0004-6361/201322698}, \href
  {https://ui.adsabs.harvard.edu/abs/2014A&A...563A..82T} {563, A82}

\bibitem[\protect\citeauthoryear{{Vink} et~al.,}{{Vink}
  et~al.}{2022}]{ixpe_casA}
{Vink} J.,  et~al., 2022, arXiv e-prints, \href
  {https://ui.adsabs.harvard.edu/abs/2022arXiv220606713V} {p. arXiv:2206.06713}

\bibitem[\protect\citeauthoryear{{Weisskopf}, {Wu}, {Tennant}, {Swartz}  \&
  {Ghosh}}{{Weisskopf} et~al.}{2004}]{weisskopf2004}
{Weisskopf} M.~C.,  {Wu} K.,  {Tennant} A.~F.,  {Swartz} D.~A.,   {Ghosh}
  K.~K.,  2004, \mn@doi [\apj] {10.1086/381307}, \href
  {https://ui.adsabs.harvard.edu/abs/2004ApJ...605..360W} {605, 360}

\bibitem[\protect\citeauthoryear{{Weisskopf} et~al.,}{{Weisskopf}
  et~al.}{2022}]{weisskopf2022}
{Weisskopf} M.~C.,  et~al., 2022, \mn@doi [Journal of Astronomical Telescopes,
  Instruments, and Systems] {10.1117/1.JATIS.8.2.026002}, \href
  {https://ui.adsabs.harvard.edu/abs/2022JATIS...8b6002W} {8, 026002}

\bibitem[\protect\citeauthoryear{{Wilson}, {Shopbell}, {Simpson},
  {Storchi-Bergmann}, {Barbosa}  \& {Ward}}{{Wilson} et~al.}{2000}]{wilson2000}
{Wilson} A.~S.,  {Shopbell} P.~L.,  {Simpson} C.,  {Storchi-Bergmann} T.,
  {Barbosa} F.~K.~B.,   {Ward} M.~J.,  2000, \mn@doi [\aj] {10.1086/301532},
  \href {https://ui.adsabs.harvard.edu/abs/2000AJ....120.1325W} {120, 1325}

\bibitem[\protect\citeauthoryear{{Yang}, {Wilson}, {Matt}, {Terashima}  \&
  {Greenhill}}{{Yang} et~al.}{2009}]{yang2009}
{Yang} Y.,  {Wilson} A.~S.,  {Matt} G.,  {Terashima} Y.,   {Greenhill} L.~J.,
  2009, \mn@doi [\apj] {10.1088/0004-637X/691/1/131}, \href
  {https://ui.adsabs.harvard.edu/abs/2009ApJ...691..131Y} {691, 131}

\makeatother
\end{thebibliography}

\section*{Appendix: $Chandra$ observations and analysis}

$Chandra$ performed two 10-ks observations  at the beginning (2022 July 11, obs1) and end (2022 July 24, obs2) of the \textit{IXPE} observation, to monitor the flux
state of the variable ULXs in the Circinus Galaxy, as well as the AGN spectrum. The \textit{Chandra} image of obs1 is shown in Fig. \ref{fig:chandra-image}. 
We analysed the spectra of CG X-1 and CG X-2 to account for these sources in the \textit{IXPE} spectral analysis. 

\begin{figure}
\includegraphics[width=1.0\columnwidth]{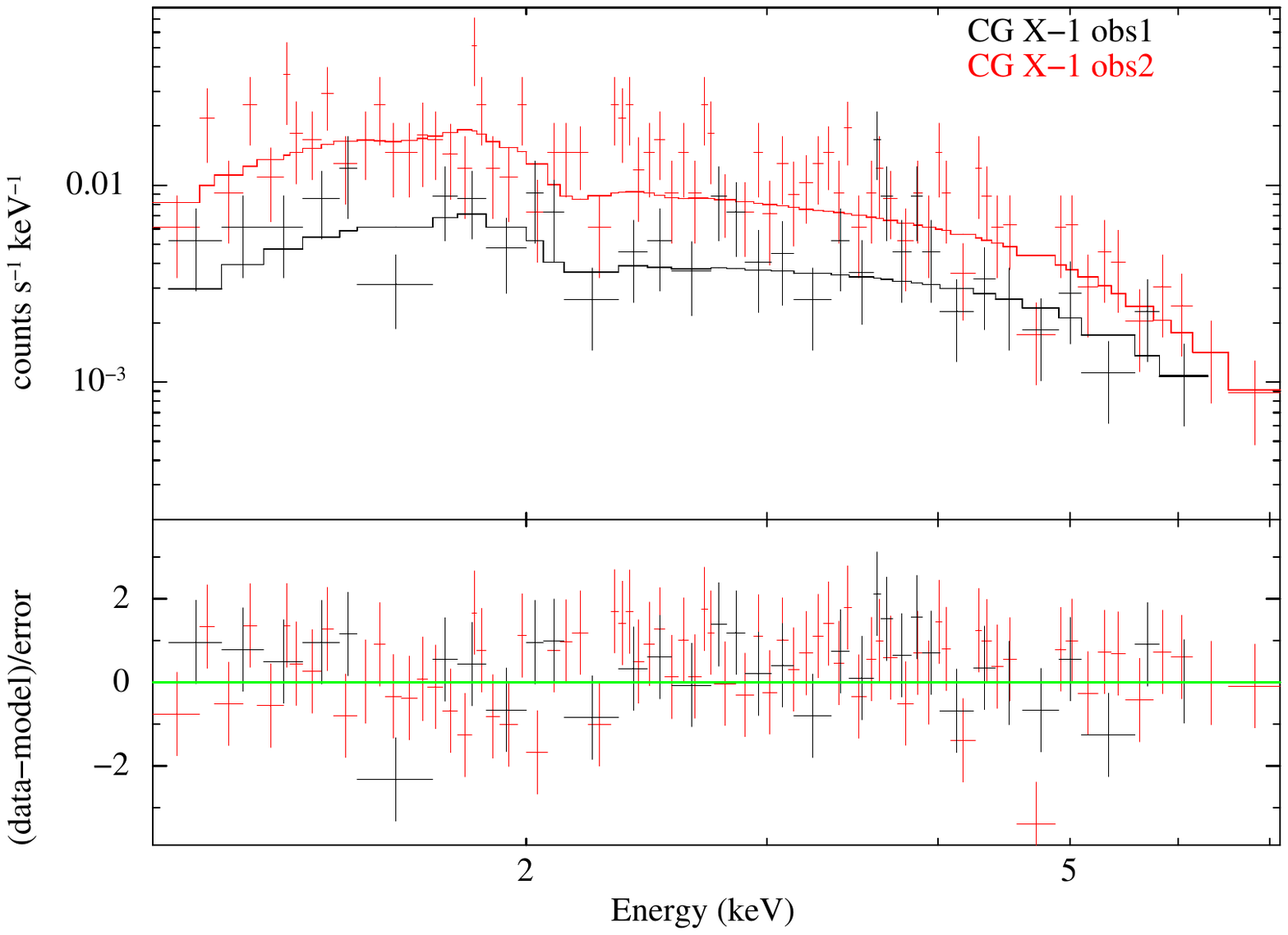}
\includegraphics[width=1.0\columnwidth]{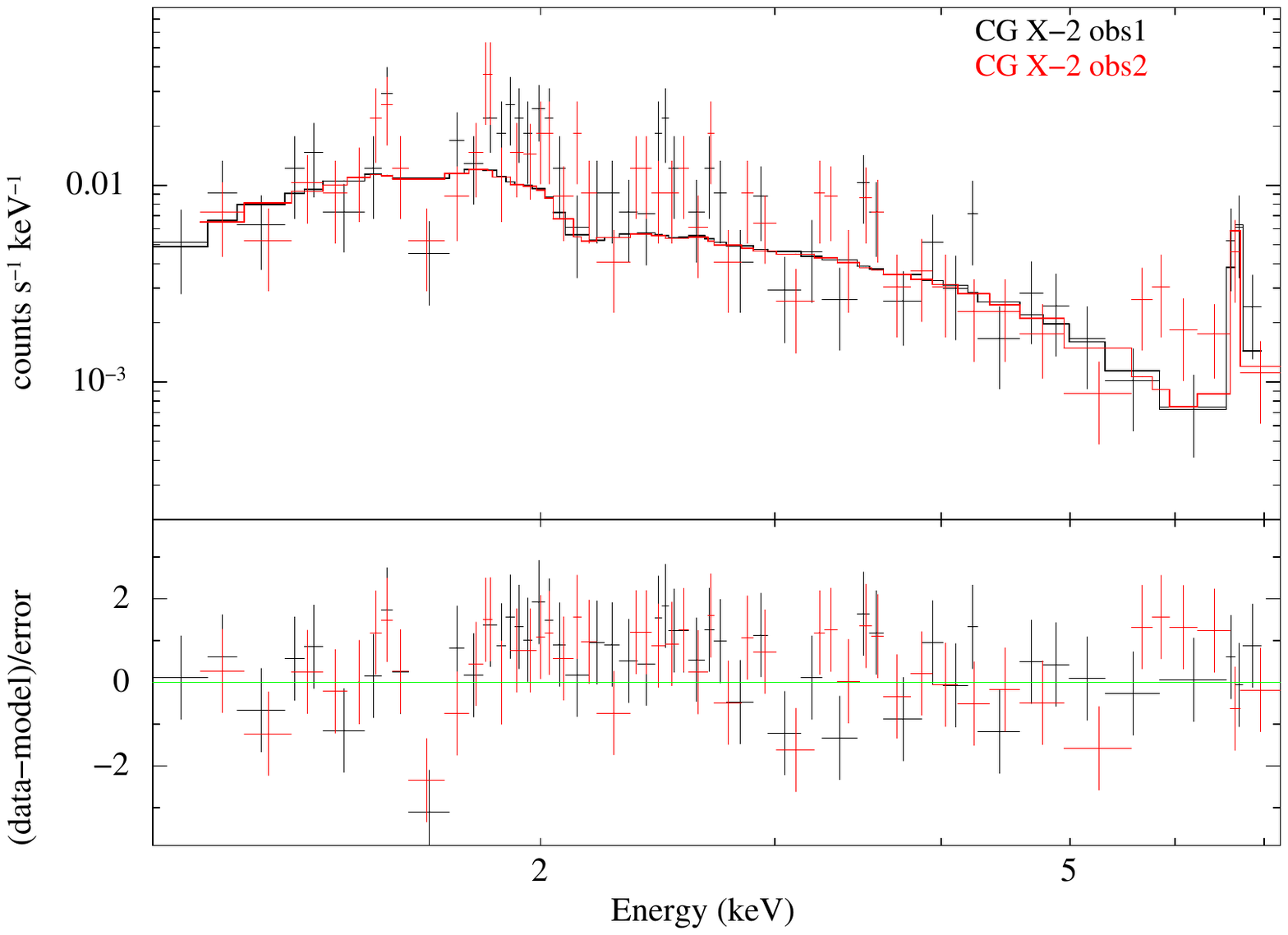}
\caption{\textit{Chandra}/ACIS spectra of CG X-1 (top) and CG X-2 (bottom) with the best-fitting model in the 1--8 keV band. The black and red coloured crosses correspond to observation 1 and 2, respectively.  
  }
\label{fig:chandra-ulx}
\end{figure}

The nature of CG X-1 has been debated in the literature \citep{bauer2001,bianchi2002,weisskopf2004,esposito2015}, however its spectral and variability properties are consistent with an eclipsing Wolf-Rayet ULX \citep{esposito2015,qiu2019}. In any case, its spectrum is well described by a variable power law. We find a good fit of the two spectra from obs1 and obs2 ($\chi^2$/d.o.f. = 109/106), with a photon index $\Gamma=0.9\pm0.2$ in obs1 and $1.30\pm0.17$ in obs2. The 1--8 keV flux increases by a factor of $\sim 2$ between the two pointings, from $(3.7\pm0.5)\times 10^{-13}$ to $(7.2\pm0.5)\times 10^{-13}$ \fluxcgs. The spectra are shown in Fig. \ref{fig:chandra-ulx} (top panel). When modeling the \textit{IXPE} spectrum, we opt to include a power law with average parameters, namely $\Gamma=1.1$ and a 1--8 keV flux of $5.45\times 10^{-13}$ \fluxcgs. 

CG X-2 is a supernova remnant candidate \citep{bauer2001}. Following \cite{bauer2001}, we fit the spectrum with a Raymond-Smith thermal plasma with a temperature of 10 keV, plus a Gaussian line for ionized iron emission \cite[see also][]{bianchi2002}. We jointly fit the two observations, as we find no significant variability among them. We find a decent fit ($\chi^2$/d.o.f. = 118/102) with a narrow Gaussian line at $6.67\pm0.05$ keV having a flux of $(8\pm4)\times 10^{-6}$ ph cm$^{-2}$ s$^{-1}$. The total model flux is $(4.8\pm0.3)\times 10^{-13}$ \fluxcgs. We show the spectra in Fig. \ref{fig:chandra-ulx} (bottom panel).

Besides \textit{Chandra}, \textit{Swift}/XRT also performed two observations of the Circinus field, centered on CG X-2, on 2022 July 18 and July 20 (ObsIds 00045807010 and 00045807011) for an exposure time of 4.3 ks and 4.6 ks, respectively. The small number of counts in XRT prevents a detailed analysis; however, we find no evidence for a significant flux variability of CG X-1 and CG X-2 during these observations. 

\begin{figure}
\includegraphics[width=1.0\columnwidth]{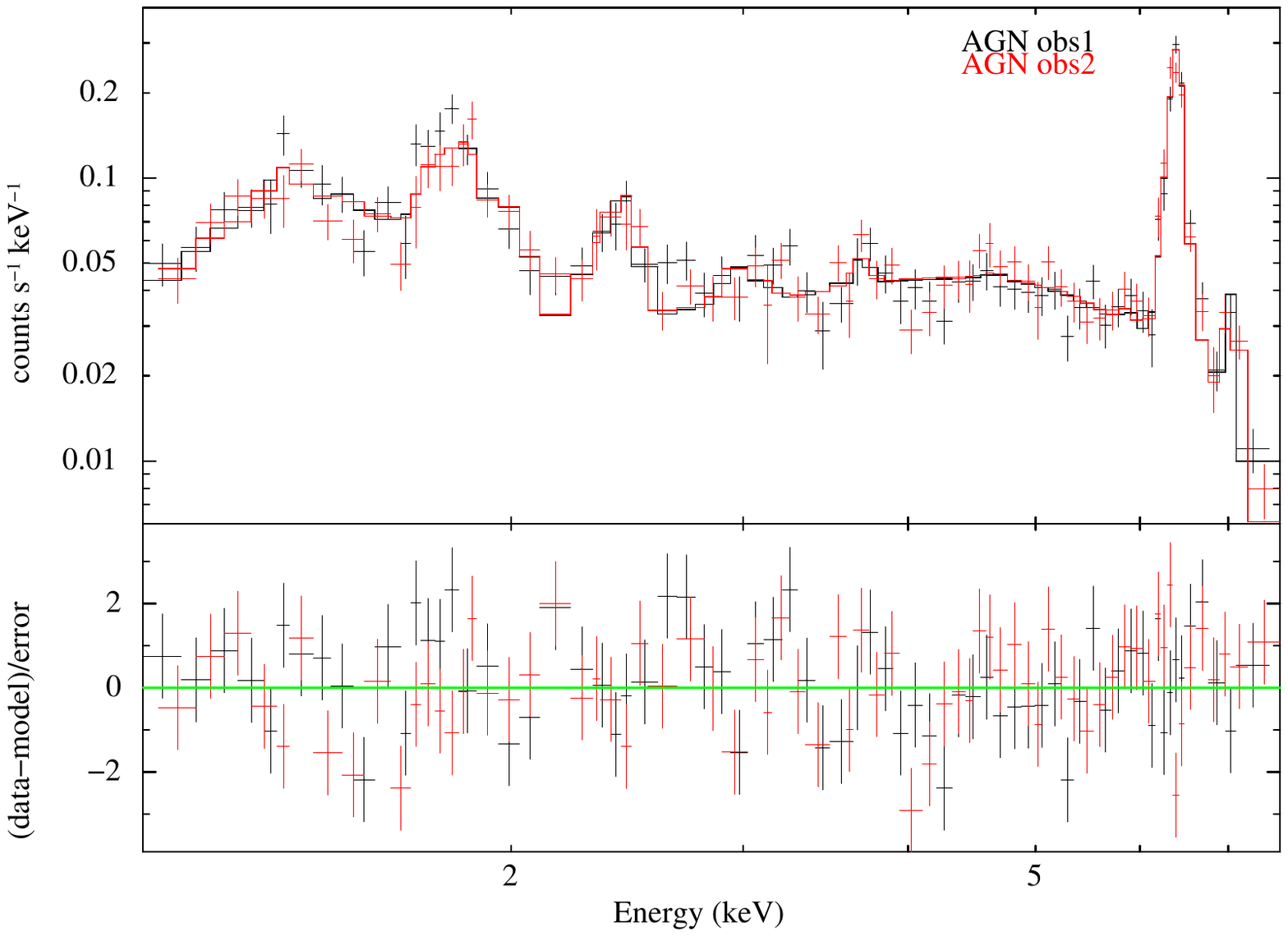}
\caption{\textit{Chandra}/ACIS spectra of the AGN with the best-fitting model in the 1--8 keV band. The black and red coloured crosses correspond to observation 1 and 2, respectively. }
\label{fig:chandra-agn}
\end{figure}

Finally, we fit the spectrum of the AGN, 
with a model including cold reflection (\textsc{pexrav}) and  Gaussian emission lines \citep{massaro2006}. As a first step, we jointly fit obs1 and obs2 keeping all parameters tied. We obtain a poor fit with $\chi^2$/d.o.f. = 340/135, with significant residuals below 2 keV. Next, we include a power law to describe warm reflection, and we fix the \textsc{pexrav} photon index at 1.6 to avoid model degeneracies. We obtain an acceptable fit with $\chi^2$/d.o.f. = 179/134, with no prominent residuals (see Fig. \ref{fig:chandra-agn}). For the warm reflection power law, we obtain a photon index of $3.0 \pm 0.2$, consistent with the value reported by \cite{marinucci2013} from \textit{Chandra} data. The flux contribution of this component is 5--10\% of the total in the 1--8 keV band, raising to $\sim$20\% in the 2--4 keV band. We do not find a significant improvement by leaving the parameters free to vary between the two observations. We also do not find a significant improvement by leaving free the \textsc{pexrav} photon index.
In the \textit{IXPE} energy band (2--8 keV), the flux of the AGN is $(1.04\pm0.02)\times 10^{-11}$ \fluxcgs, while that of CG X-1 and CG X-2 is respectively 3--6 and 4 per cent of the AGN flux. 

\section*{Authors' affiliations}
% List of institutions
\newcounter{foo}  
\begin{list}{$^{\arabic{foo}}$}
    {\usecounter{foo}
     \setlength{\labelwidth}{0em}
     \setlength{\labelsep}{0em}
     \setlength{\itemsep}{0pt}
     \setlength{\leftmargin}{0cm}
     \setlength{\rightmargin}{0cm}
     \setlength{\itemindent}{0em} 
    }
\item Dipartimento di Matematica e Fisica, Universit\`a degli Studi Roma Tre, via della Vasca Navale 84, 00146 Roma, Italy \label{aff:UniRoma3}
\item ASI - Agenzia Spaziale Italiana, Via del Politecnico snc, 00133 Roma, Italy \label{aff:ASI}
\item Université de Strasbourg, CNRS, Observatoire Astronomique de Strasbourg, UMR 7550, 67000 Strasbourg, France \label{aff:Strasbourg}
\item Kavli Institute for Astrophysics and Space Research, Massachusetts Institute of Technology, 77 Massachusetts Ave., Cambridge, MA 02139, USA \label{aff:MIT}
\item Department of Physics and Astronomy, 20014 University of Turku, Finland \label{aff:Turku}
\item Space Research Institute of the Russian Academy of Sciences, Profsoyuznaya Str. 84/32, Moscow 117997, Russia \label{aff:IKI}
\item INAF Istituto di Astrofisica e Planetologia Spaziali, Via del Fosso del Cavaliere 100, 00133 Roma, Italy \label{aff:INAF-IAPS}
\item California Institute of Technology, Pasadena, CA 91125, USA \label{aff:Caltech}
\item Department of Physics - Astrophysics, University of Oxford, Denys Wilkinson Building, Keble Road, Oxford OX1 3RH, UK \label{aff:Oxford}
\item School of Mathematics, Statistics and Physics, Newcastle University, Herschel Building, Newcastle upon Tyne,\:NE1\,7RU, UK \label{aff:Newcastle}
\item Dipartimento di Fisica, Università degli Studi di Roma "La Sapienza", Piazzale Aldo Moro 5, 00185 Roma, Italy \label{aff:UniRoma1}
\item Dipartimento di Fisica, Università degli Studi di Roma "Tor Vergata", Via della Ricerca Scientifica 1, 00133 Roma, Italy \label{aff:UniRoma2}
\item Physics Department and McDonnell Center for the Space Sciences, Washington University in St. Louis, St. Louis, MO 63130, USA \label{aff:WUStL}
\item Astronomical Institute of the Czech Academy of Sciences, Boční II 1401/1, 14100 Praha 4, Czech Republic \label{aff:CAS-ASU}
\item Istituto Nazionale di Fisica Nucleare, Sezione di Roma "Tor Vergata", Via della Ricerca Scientifica 1, 00133 Roma, Italy \label{aff:INFN-Roma2}
\item Department of Astronomy, University of Maryland, College Park, Maryland 20742, USA \label{aff:UMd}
\item NASA Marshall Space Flight Center,Huntsville,AL 35812,USA \label{aff:NASA-MSFC}
\item Astronomical Institute, Charles University, V Holešovičkách 2, CZ-18000 Prague, Czech Republic \label{aff:Charles}
\item Instituto de Astrofísica de Andalucía—CSIC, Glorieta de la Astronomía s/n, 18008 Granada, Spain \label{aff:CSIC-IAA}
\item INAF Osservatorio Astronomico di Roma, Via Frascati 33, 00078 Monte Porzio Catone (RM), Italy \label{aff:INAF-OAR}
\item Space Science Data Center, Agenzia Spaziale Italiana, Via del Politecnico snc, 00133 Roma, Italy \label{aff:ASI-SSDC}
\item INAF Osservatorio Astronomico di Cagliari, Via della Scienza 5, 09047 Selargius (CA), Italy \label{aff:INAF-OAC}
\item Istituto Nazionale di Fisica Nucleare, Sezione di Pisa, Largo B. Pontecorvo 3, 56127 Pisa, Italy \label{aff:INFN-PI}
\item Dipartimento di Fisica, Università di Pisa, Largo B. Pontecorvo 3, 56127 Pisa, Italy \label{aff:UniPI}
\item Istituto Nazionale di Fisica Nucleare, Sezione di Torino, Via Pietro Giuria 1, 10125 Torino, Italy \label{aff:INFN-TO}
\item Dipartimento di Fisica, Università degli Studi di Torino, Via Pietro Giuria 1, 10125 Torino, Italy \label{aff:UniTO}
\item INAF Osservatorio Astrofisico di Arcetri, Largo Enrico Fermi 5, 50125 Firenze, Italy \label{aff:INAF-FI}
\item Dipartimento di Fisica e Astronomia, Università degli Studi di Firenze, Via Sansone 1, 50019 Sesto Fiorentino (FI), Italy \label{aff:UniFI}
\item Istituto Nazionale di Fisica Nucleare, Sezione di Firenze, Via Sansone 1, 50019 Sesto Fiorentino (FI), Italy \label{aff:INFN-FI}
\item Department of Physics and Kavli Institute for Particle Astrophysics and Cosmology, Stanford University, Stanford, California 94305, USA \label{aff:Stanford}
\item Institut f\"ur Astronomie und Astrophysik, Universit\"at T\"ubingen, Sand 1, 72076 T\"ubingen, Germany \label{aff:Tuebingen}
\item RIKEN Cluster for Pioneering Research, 2-1 Hirosawa, Wako, Saitama 351-0198, Japan \label{aff:RIKEN}
\item Yamagata University, 1-4-12 Kojirakawa-machi, Yamagata-shi 990-8560, Japan \label{aff:yamagata} 
\item University of British Columbia,Vancouver,BC V6T 1Z4,Canada \label{aff:UBC} 
\item Department of Physics, Faculty of Science and Engineering, Chuo University, 1-13-27 Kasuga, Bunkyo-ku, Tokyo 112-8551, Japan \label{aff:Chuo}
\item Institute for Astrophysical Research, Boston University, 725 Commonwealth Avenue, Boston, MA 02215, USA \label{aff:BU} 
\item Department of Astrophysics, St. Petersburg State University, Universitetsky pr. 28, Petrodvoretz, 198504 St. Petersburg, Russia \label{aff:SBU}
\item Finnish Centre for Astronomy with ESO, 20014 University of Turku, Finland \label{aff:FINCA}
\item Graduate School of Science, Division of Particle and Astrophysical Science, Nagoya University, Furo-cho, Chikusa-ku, Nagoya, Aichi 464-8602, Japan \label{aff:nagoya}
\item Hiroshima Astrophysical Science Center, Hiroshima University, 1-3-1 Kagamiyama, Higashi-Hiroshima, Hiroshima 739-8526, Japan \label{aff:hiroshima}
\item Department of Physics, The University of Hong Kong, Pokfulam, Hong Kong \label{aff:HKU}
\item Department of Astronomy and Astrophysics, Pennsylvania State University, University Park, PA 16802, USA \label{aff:PSU}
\item Université Grenoble Alpes,CNRS,IPAG,38000 Grenoble,France \label{aff:Grenoble}
\item Center for Astrophysics | Harvard \& Smithsonian, 60 Garden St, Cambridge, MA 02138, USA \label{aff:CfA}
\item INAF Osservatorio Astronomico di Brera, Via E. Bianchi 46, 23807 Merate (LC), Italy \label{aff:INAF-OAB}
\item Dipartimento di Fisica e Astronomia, Università degli Studi di Padova, Via Marzolo 8, 35131 Padova, Italy \label{aff:UniPD}
\item Mullard Space Science Laboratory, University College London, Holmbury St Mary, Dorking, Surrey RH5 6NT, UK \label{aff:MSSL}
\item Anton Pannekoek Institute for Astronomy \& GRAPPA, University of Amsterdam, Science Park 904, 1098 XH Amsterdam, The Netherlands \label{aff:amsterdam}
\item Guangxi Key Laboratory for Relativistic Astrophysics, School of Physical Science and Technology, Guangxi University, Nanning 530004, China \label{aff:GXU}
\end{list}

% Don't change these lines
\bsp	% typesetting comment
\label{lastpage}

\end{document}